\documentclass[11pt,draftclsnofoot,onecolumn]{IEEEtran}

\usepackage[usenames]{color}
 \usepackage{verbatim}

\usepackage{latexsym}
\usepackage[cmex10]{amsmath}
\usepackage{amstext,amssymb, amsfonts, amsthm}

\usepackage{graphicx,wrapfig}

\usepackage{wrapfig}

\def\mindex#1{\index{#1}}

\def\sq{\hbox{\rlap{$\sqcap$}$\sqcup$}}
\def\qed{\ifmmode\sq\else{\unskip\nobreak\hfil
\penalty50\hskip1em\null\nobreak\hfil\sq
\parfillskip=0pt\finalhyphendemerits=0\endgraf}\fi\medskip}

\long\def\defbox#1{\framebox[.9\hsize][c]{\parbox{.85\hsize}{%
\parindent=0pt
\baselineskip=12pt plus .1pt      
\parskip=6pt plus 1.5pt minus 1pt 
 #1}}}

\long\def\beginbox#1\endbox{\subsection*{}%
\hbox{\hspace{.05\hsize}\defbox{\medskip#1\bigskip}}%
\subsection*{}}

\def\endbox{}

\def\transpose{{\hbox{\it\tiny T}}}

\newsavebox{\junk}
\savebox{\junk}[1.6mm]{\hbox{$|\!|\!|$}}

\def\liminf{\mathop{\rm lim\ inf}}
\def\argmin{\mathop{\rm arg\, min}}

\def\argmax{\mathop{\rm arg\, max}}

\def\state{{\sf X}}

\def\zstate{{\sf Z}}

\newcommand{\field}[1]{\mathbb{#1}}

\def\Re{\field{R}}

\def\ind{\field{I}}

\def\cpi{\check{\pi}}

\def\bfmath#1{{\mathchoice{\mbox{\boldmath$#1$}}%
{\mbox{\boldmath$#1$}}%
{\mbox{\boldmath$\scriptstyle#1$}}%
{\mbox{\boldmath$\scriptscriptstyle#1$}}}}

\def\bfmX{\bfmath{X}}

\def\bfmY{\bfmath{Y}}

\def\bfmhhaY{\bfmath{\hhaY}} 
\def\bfmhhaY{\hbox to 0pt{$\widehat{\bfmY}$\hss}\widehat{\phantom{\raise 1.25pt\hbox{$\bfmY$}}}}

\def\bfmW{\bfmath{W}}  
  
\def\bfmZ{\bfmath{Z}}

\def\bfphi{\bfmath{\phi}}

\def\til={{\widetilde =}}

\def\clB{{\cal B}}

\def\clE{{\cal E}}
\def\clF{{\cal F}}
\def\clG{{\cal G}}
\def\clH{{\cal H}}

\def\clN{{\cal N}}

\def\clP{{\cal P}}
\def\clQ{{\cal Q}}

\def\clT{{\cal T}}

\def\Var{\hbox{\sf Var}\,}

\def\atop#1#2{\genfrac{}{}{0pt}{}{#1}{#2}}

 \def\FRAC#1#2#3{\genfrac{}{}{}{#1}{#2}{#3}}

\def\ddtp{{\mathchoice{\FRAC{1}{d^{\hbox to 2pt{\rm\tiny +\hss}}}{dt}}%
{\FRAC{1}{d^{\hbox to 2pt{\rm\tiny +\hss}}}{dt}}%
{\FRAC{3}{d^{\hbox to 2pt{\rm\tiny +\hss}}}{dt}}%
{\FRAC{3}{d^{\hbox to 2pt{\rm\tiny +\hss}}}{dt}}}}

\def\half{{\mathchoice{\FRAC{1}{1}{2}}%
{\FRAC{1}{1}{2}}%
{\FRAC{3}{1}{2}}%
{\FRAC{3}{1}{2}}}}

\def\fourth{{\mathchoice{\FRAC{1}{1}{4}}%
{\FRAC{1}{1}{4}}%
{\FRAC{3}{1}{4}}%
{\FRAC{3}{1}{4}}}}

\def\eqdef{\mathbin{:=}}

\def\Prob{{\sf P}}

\def\Expect{{\sf E}}

\def\average#1,#2,{{1\over #2} \sum_{#1}^{#2}}

\def\eye(#1){{\bf(#1)}\quad}

\newtheorem{theorem}{Theorem}[section]
\newtheorem{corollary}{Corollary}[section]
\newtheorem{proposition}[theorem]{Proposition}
\newtheorem{lemma}[theorem]{Lemma}

\def\Lemma#1{Lemma~\ref{t:#1}}
\def\Proposition#1{Proposition~\ref{t:#1}}
\def\Theorem#1{Theorem~\ref{t:#1}}
\def\Corollary#1{Corollary~\ref{t:#1}}

\def\Section#1{Section~\ref{#1}}

\def\Figure#1{Figure~\ref{f:#1}}

\def\eq#1/{(\ref{e:#1})}

\newcommand{\beqn}[1]{\notes{#1}%
\begin{eqnarray} \elabel{#1}}

\newcommand{\eeqn}{\end{eqnarray} }

\newcommand{\beq}[1]{\notes{#1}%
\begin{equation}\elabel{#1}}

\newcommand{\eeq}{\end{equation}}

\def\bdes{\begin{description}}
\def\edes{\end{description}}

\newcommand{\barx}{{\bar{x}}}

\newcounter{rmnum}
\newenvironment{romannum}{\begin{list}{{\upshape (\roman{rmnum})}}{\usecounter{rmnum}
\setlength{\leftmargin}{14pt}
\setlength{\rightmargin}{8pt}
\setlength{\itemindent}{-1pt}
}}{\end{list}}

\newcounter{anum}
\newenvironment{alphanum}{\begin{list}{{\upshape (\alph{anum})}}{\usecounter{anum}
\setlength{\leftmargin}{14pt}
\setlength{\rightmargin}{8pt}
\setlength{\itemindent}{-1pt}
}}{\end{list}}

{\end{list}}

\def\ass(#1:#2){(#1\ref{#1:#2})}

\def\ritem#1{
\item[{\sf \ass(\current_model:#1)}]
}

\newenvironment{recall-ass}[1]{%
\begin{description}
\def\current_model{#1}}{
\end{description}
}

\def\Ebox#1#2{%
\begin{center}
 \parbox{#1\hsize}{\epsfxsize=\hsize \epsfbox{#2}}
\end{center}}

\newcommand{\bd}{\begin{description}}
\newcommand{\ed}{\end{description}}
\newcommand{\bt}{\begin{theorem}}
\newcommand{\et}{\end{theorem}}
\newcommand{\ba}{\begin{array}{rcl}}
\newcommand{\ea}{\end{array}}

\def\breakMed{\\[.2cm]}

\newlength{\noteWidth}
\long\def\notes#1{\ifinner
             {\tiny #1}
             \else
              \marginpar{\parbox[t]{\noteWidth}{\raggedright\tiny #1}}
               \fi}

\def\notes#1{\typeout{#1 !!!}}  

 \def\Ebox#1#2{%
 \begin{center}
\includegraphics[width=#1\hsize]{#2}
 \end{center}
\vspace{-.45cm}}

\def\barx{\bar{x}}

\def\tilV{\widetilde{V}}
\def\tilY{\widetilde{Y}}
\def\tilS{\widetilde{S}}
\def\har{\hat r}
\def\bP{\mathbb{P}}

\newcounter{ass}
\newenvironment{assumption}{\begin{list}{{\upshape \bf (A\arabic{ass}) \ }}{\usecounter{ass}
\setlength{\leftmargin}{14pt}
\setlength{\rightmargin}{12pt}
\setlength{\itemindent}{-1pt}
}}{\end{list}}

\def\Ass#1{(A\ref{ass:#1})}

 \def\pz{\clP(\zstate)}

\def\tH{{\hbox{\rm\tiny H}}}
\def\tG{{\hbox{\rm\tiny GLRT}}}

\def\DMM{D^{\hbox{\rm\tiny MM}}}
\def\DR{D^{\hbox{\rm\tiny ROB}}}
\def\QMM{\clQ^{\hbox{\rm\tiny MM}}}

\def\MM{{\hbox{\rm\tiny MM}}}
\def\FA{{\hbox{\rm\tiny FA}}}
\def\HLLR{\clH^{\hbox{\rm\tiny LLR}}}

\def\uni{{\hbox{\rm\tiny u}}}

\def\trace{{\sf trace}}
\def\Cov{{\sf Cov}}
\def\Expect{{\sf E}}
\begin{document}

\title{\null
Universal and Composite Hypothesis Testing\\ via Mismatched Divergence}
\author{Jayakrishnan Unnikrishnan, Dayu Huang, \\Sean
Meyn, Amit Surana and Venugopal Veeravalli\footnote[1]{
Amit Surana is with United Technologies Research Center, 411
Silver Lane, E.~Hartford, CT. Email: SuranaA@utrc.utc.com. The
remaining authors are with the Department of Electrical and
Computer Engineering and the Coordinated Science Laboratory,
University of Illinois at Urbana-Champaign, Urbana, IL. Email:
\{junnikr2, dhuang8, meyn, vvv\}@illinois.edu.

This research was
partially supported by NSF under grant CCF 07-29031 and by
UTRC. Any opinions, findings, and conclusions or
recommendations expressed in this material are those of the
authors and do not necessarily reflect the views of the  NSF or
UTRC.

Portions of the results presented here were published in abridged form in \cite{huaunnmeyveesur09}.
}}
\maketitle


\begin{abstract}
For the \textit{universal} hypothesis testing problem, where
the goal is to decide between the known null hypothesis
distribution and some other unknown distribution, Hoeffding
proposed a universal test in the nineteen sixties. Hoeffding's
universal test statistic can be written in terms of
Kullback-Leibler (K-L) divergence between the empirical
distribution of the observations and the null hypothesis
distribution. In this paper a modification of Hoeffding's test
is considered based on a relaxation of the K-L divergence, referred to as the mismatched divergence. The
resulting mismatched test is shown to be a generalized
likelihood-ratio test (GLRT) for the case where the alternate
distribution lies in a parametric family of distributions
characterized by a finite dimensional parameter, i.e., it is a
solution to the corresponding \textit{composite} hypothesis
testing problem. For certain choices of the alternate
distribution, it is shown that both the Hoeffding test and the
mismatched test have the same asymptotic performance in terms
of error exponents. A consequence of this result is that the
GLRT is optimal in differentiating a particular distribution
from others in an exponential family. It is also shown that the
mismatched test has a significant advantage over the Hoeffding
test in terms of finite sample size performance for applications involving large alphabet distributions. This advantage
is due to the difference in the asymptotic variances of the two
test statistics under the null hypothesis.




 \smallskip


\end{abstract}


\thispagestyle{empty}


\section{Introduction and Background}

This paper is concerned with the following hypothesis testing
problem:  Suppose that the observations $\bfmZ = \{Z_t :
t=1,\ldots \}$ form an i.i.d.\ sequence evolving on a set of
cardinality $N$, denoted by $\zstate = \{z_1, z_2, \ldots,
z_N\}$.       Based on observations of this sequence we wish to
decide if the marginal distribution of the observations is a
given distribution $\pi^0$, or some other distribution
{$\pi^1$} that is either unknown or known only to belong to a
certain class of distributions. {When the observations have
distribution $\pi^0$ we say that the \textit{null hypothesis}
is true, and when the observations have some other distribution
$\pi^1$ we say that the \textit{alternate hypothesis} is true.}

A decision rule is characterized by a \textit{sequence} of
tests $\bfphi\eqdef \{\phi_n : n\ge 1\}$, where  $\phi_n:
\zstate^n \mapsto \{0, 1\}$ with $\zstate^n$ representing the
$n$-th order Cartesian-product of $\zstate$. The decision based
on the first $n$ elements of the observation sequence is given
by $\phi_n(Z_1, Z_2, \ldots, Z_n)$, where $\phi_n =0$
represents a decision in favor of accepting $\pi^0$ as the true
marginal distribution.

The set of probability measures on $\zstate$ is denoted $\pz$.
The relative entropy (or Kullback-Leibler divergence) between
two distributions $\nu^1, \nu^2 \in \pz$ is denoted $D(\nu^1 \|
\nu^2)$,  and for a given $\mu \in \pz$ and $\eta>0$ the
\textit{divergence ball} of \textit{radius} $\eta$ around $\mu$
is defined as,
\begin{equation}
\clQ_\eta(\mu) \eqdef \{\nu \in \pz: D(\nu \| \mu) < \eta\}. \label{e:Qball}
\end{equation}
The empirical distribution or \textit{type} of the finite set
of observations $(Z_1, Z_2, \ldots, Z_n)$ is a random variable
$\Gamma^n$ taking values in $ \pz$:
\begin{equation}
\Gamma^n(z) = \frac{1}{n} \sum_{i=1}^n \ind\{Z_i =z\},\qquad z\in\zstate
  \label{eqn:typeiid}
\end{equation}
where $\ind$ denotes the indicator function.

In the general universal hypothesis testing problem, {the null
distribution $\pi^0$ is known exactly, but no prior information
is available regarding the alternate distribution $\pi^1$.
Hoeffding proposed in \cite{hoe65a} a generalized
likelihood-ratio test (GLRT) for the universal hypothesis
testing problem, in which the alternate distribution $\pi^1$ is
unrestricted  --- it is an arbitrary distribution in $\pz$, the
set of probability distributions on $\zstate$. Hoeffding's test
sequence is given by, }
\begin{equation}
\begin{aligned}
\phi_n^\tH &= \ind\{\sup_{\pi^1 \in \pz} \frac{1}{n} \sum_{i=1}^n \log \frac{\pi^1(Z_i)}{\pi^0(Z_i)} \geq \eta \}
\label{eqn:Hoefftestnew}
\end{aligned}
\end{equation}
It is easy to see that the \textit{Hoeffding test} (\ref{eqn:Hoefftestnew}) can be rewritten as follows:
\begin{equation}
\begin{aligned}
\phi_n^\tH &= \ind\{\frac{1}{n} \sum_{i=1}^n \log \frac{\Gamma^n(Z_i)}{\pi^0(Z_i)} \geq \eta \}
\breakMed
&= \ind\{ \sum_{z\in\zstate} \Gamma^n(z)\log \frac{\Gamma^n(z)}{\pi^0(z)} \geq \eta \}
\breakMed
&= \ind\{ D( \Gamma^n \| \pi^0) \geq \eta \}
\breakMed
&= \ind\{\Gamma^n \notin \clQ_\eta(\pi^0) \}
\label{eqn:Hoefftestnew2}
\end{aligned}
\end{equation}

If we have some prior information on the alternate distribution
{$\pi^1$}, a different version of the GLRT is used. In
particular, suppose it is known that the alternate distribution
lies in a parametric family of distributions of the following
form:
\[
\clE_{\pi^0} \eqdef \{\cpi^r : r \in \Re ^d \}.
\]
where $\cpi^r \in \pz$ are probability distributions on
$\zstate$ parameterized by a parameter $r \in \Re^d$. The
specific form of $\cpi^r$ is defined later in the paper. In
this case, the resulting composite hypothesis testing problem
is typically solved using a GLRT (see \cite{zeizivmer92} for
results related to the present paper, and \cite{levmer02} for a
more recent account) of the following form:
\begin{equation}
\begin{aligned}
\phi_n^\MM
   &= \ind\{\sup_{\pi^1 \in \clE_{\pi^0}} \sum_{z \in \zstate} \Gamma^n(z)
\log\frac{\pi^1(z)}{\pi^0(z)} \geq \eta \} \,.
\label{eqn:MMtestnew}
\end{aligned}
\end{equation}
We show that this test can be interpreted as a relaxation of
the Hoeffding test of (\ref{eqn:Hoefftestnew2}). {In particular
we show that \eqref{eqn:MMtestnew} can be expressed in a form
similar to \eqref{eqn:Hoefftestnew2},
\begin{equation}
\begin{aligned}
\phi_n^\MM &= \ind\{ \DMM( \Gamma^n \| \pi^0) \geq \eta \}
\label{eqn:MMtestnew2}
\end{aligned}
\end{equation}
where $\DMM$ is  the \textit{mismatched divergence};
 a relaxation of the K-L divergence, in the sense that $ \DMM( \mu \| \pi)\le D( \mu \| \pi)$ for any $\mu,\pi\in\pz$.
We refer to the test (\ref{eqn:MMtestnew2}) as the \textit{mismatched test}.

This paper is devoted to the analysis of   the mismatched divergence and mismatched  test.




The terminology is borrowed from the \textit{mismatched
channel} (see Lapidoth \cite{lap96} for a bibliography). } The
mismatched divergence described here is a generalization of the
relaxation introduced in \cite{abbmedmeyzhe07a}. In this way we
embed the analysis of the resulting universal test within the
framework of Csisz\'{a}r and Shields \cite{csishi04}. The
mismatched test statistic can also be viewed as a
generalization of the robust hypothesis testing statistic
introduced in \cite{panmeyvee04a,panmey06a}.

When the alternate distribution satisfies $\pi^1 \in
\clE_{\pi^0}$, we  show that, under some regularity conditions
on $\clE_{\pi^0}$, the mismatched test of
(\ref{eqn:MMtestnew2}) and Hoeffding's test of
(\ref{eqn:Hoefftestnew2}) have identical asymptotic performance
in terms of error exponents. A consequence of this result is
that the GLRT is optimal in differentiating a particular
distribution from others in an exponential family of
distributions. We also establish that the proposed mismatched
test has a significant advantage over the Hoeffding test in
terms of finite sample size performance. This advantage is due
to the difference in the asymptotic variances of the two test
statistics under the null hypothesis. In particular, we show
that the variance of the K-L divergence grows linearly with the
alphabet size, making the test impractical for applications
involving large alphabet distributions. We also show that the
variance of the mismatched divergence  grows linearly with the
dimension $d$ of the parameter space, and can hence be
controlled through a prudent choice of the function class
defining the mismatched divergence.

\medskip

The remainder of the paper is organized as follows.   We begin
in \Section{s:mm} with a description of mismatched divergence
and the mismatched test,   and describe their relation to other
concepts including robust hypothesis testing, composite
hypothesis testing, reverse I-projection, and maximum
likelihood (ML) estimation. Formulae for the asymptotic mean
and variance of the test statistics are presented in
\Section{s:var}. \Section{s:var} also contains a discussion
interpreting these asymptotic results in terms of the
performance of the detection rule. Proofs of the main results
are provided in the appendix. Conclusions and directions for
future research are contained in \Section{s:conc}.

%
%
%
%
%
%
%
%

\section{Mismatched Divergence}
\label{s:mm}

We adopt the following compact notation in the paper:  For any
function $f\colon\zstate\to\Re$ and $\pi\in\pz$ we denote the
mean $\sum_{z \in \zstate} f(z)\pi(z)$ by $\pi(f)$, or by
$\langle \pi,f\rangle$ when we wish to emphasize the
convex-analytic setting.   At times we will extend these
definitions to allow functions $f$ taking values in a vector
space. For $z \in \zstate$ and $\pi\in\pz$, we still use
$\pi(z)$ to denote the probability assigned to element $z$
under measure $\pi$. The meaning of such notation will be clear
from context.

The logarithmic moment generating function (log-MGF)
is denoted
\[
\Lambda_\pi(f) = \log (\pi(\exp(f)))
\]
where $\pi(\exp(f)) = \sum_{z \in \zstate} \pi(z) \exp(f(z))$
by the notation we introduced in the previous paragraph.  For
any two probability measures $\nu^1, \nu^2 \in \pz$ the
relative entropy is expressed,
\begin{eqnarray*}
D(\nu^1 \| \nu^2) =\left\{ \begin{tabular}{cc}
$\langle \nu^1, \log(\nu^1/\nu^2) \rangle$& if $\nu^1 \prec \nu^2$\\
$\infty$ &else
\end{tabular} \right.
\end{eqnarray*}
where   $\nu^1 \prec \nu^2$ denotes absolute continuity. The
following proposition recalls a well-known variational
representation. This can be obtained, for instance, by
specializing the representation in \cite{donvarI-II} to an
i.i.d. setting. An alternate variational representation of the
divergence is introduced in \cite{nguwaijor08}.

\begin{proposition}
\label{t:DivRep}
The relative entropy can be expressed as
the convex dual of the log moment generating function:
For any two probability measures  $\nu^1,\nu^2\in \pz$,
\begin{equation}
D(\nu^1\| \nu^2) = \sup_{f}  \bigl(\nu^1(f) - \Lambda_{\nu^2}(f) \bigr)
\label{eqn:DivRep}
\end{equation}
where the supremum is taken over the space of all real-valued
functions on $\zstate$. Furthermore, if $\nu^1$ and $\nu^2$
have equal supports, then the supremum is achieved by the log
likelihood ratio function $f^* = \log (\nu^1/\nu^2)$.
\end{proposition}

\begin{IEEEproof}
[Outline of proof] Although the result is well known, we
provide a simple proof here since similar arguments will be
reused later in the paper.

For any function $f$ and probability measure $\nu$ we have,
\begin{eqnarray*}
D(\nu^1\| \nu^2) &=& \langle \nu^1, \log(\nu^1/\nu^2) \rangle\\
               &=& \langle \nu^1, \log(\nu/\nu^2) \rangle  + \langle \nu^1, \log(\nu^1/\nu) \rangle
\end{eqnarray*}
On setting $\nu = \nu^2 \exp(f-\Lambda_{\nu^2}(f))$ this gives,
\[
D(\nu^1\| \nu^2) = \nu^1(f)  - \Lambda_{\nu^2}(f)  +  D(\nu^1\|\nu)  \ge \nu^1(f)  - \Lambda_{\nu^2}(f).
\]
If $\nu^1$ and $\nu^2$ have equal supports, then the above
inequality holds with equality for $f = \log (\nu^1/\nu^2)$,
which would lead to $\nu =\nu^1$. This proves that
\eqref{eqn:DivRep} holds whenever $\nu^1$ and $\nu^2$ have
equal supports. The proof for general distributions is similar
and is omitted here.
\end{IEEEproof}
The representation \eqref{eqn:DivRep} is the basis of the
mismatched divergence. We fix a set of functions denoted by
$\clF$, and obtain a lower bound on the relative entropy by
taking the supremum over the smaller set as follows,
\begin{equation}
\DMM(\nu^1\| \nu^2) \eqdef \sup_{f\in\clF}  \bigl\{\nu^1(f) - \Lambda_{\nu^2}(f) \bigr\}.
\label{eqn:DivMMclF}
\end{equation}
If $\nu^1$ and $\nu^2$ have full support, and if the function
class $\clF$ contains the log-likelihood ratio function $f^*
=\log( \nu^1/\nu^2)$, then it is immediate from
\Proposition{DivRep} that the supremum in (\ref{eqn:DivMMclF})
is achieved by $f^*$, and in this case $\DMM(\nu^1\| \nu^2) =
D(\nu^1\| \nu^2)$. Moreover, since the objective function in
(\ref{eqn:DivMMclF}) is invariant to shifts of $f$, it follows
that even if a constant scalar is added to the function $f^*$,
it still achieves the supremum in (\ref{eqn:DivMMclF}).

In this paper the function class is assumed to be defined
through a finite-dimensional parametrization  of the form,
\begin{equation}
\clF =\{ f_r : r\in\Re^d\}
\label{e:clF}
\end{equation}
Further assumptions will be imposed in our main results.  In
particular,  we will assume that $f_r(z)$ is differentiable as
a   function of $r$ for each $z$.

We fix a  distribution $\pi \in \pz$ and a function class of
the form \eqref{e:clF}.    For each $r \in \Re^d$ the
\textit{twisted distribution}  $\cpi^r \in \pz$ is defined as,
    \notes{8-16 -- Isn't this always $\pi^0$, not $\pi$?}
\begin{equation}
\cpi^r \eqdef \pi \exp(f_r - \Lambda_\pi(f_r) ) .
\label{eqn:twisteddistrn}
\end{equation}
The collection of all such distributions parameterized by $r$
is denoted
\begin{equation}
\clE_\pi \eqdef \{\cpi^r: r \in \Re ^d \}.\label{eqn:family}
\end{equation}

\subsection{Applications}

The applications of mismatched divergence include those
applications surveyed in Section~3 of \cite{levmer02} in their
treatment of generalized likelihood ratio tests.     Here we
list potential applications in three domains: Hypothesis
testing, source coding, and nonlinear filtering.   Other
applications include channel coding and signal detection,
following \cite{levmer02}.

\subsubsection{Hypothesis testing}
The problem of universal hypothesis testing is relevant in
several practical applications including anomaly detection. It
is often possible to have an accurate model of the normal
behavior of a system, which is usually represented by the null
hypothesis distribution $\pi^0$. The anomalous behavior is
often unknown, which is represented by the unknown alternate
distribution. The primary motivation for our research is  to
improve the {finite sample size} performance of Hoeffding's
universal hypothesis test \eqref{eqn:Hoefftestnew}.  The
difficulty we address is the large variance of this test
statistic when the alphabet size is large.   \Theorem{biasvar1}
makes this precise:

\begin{theorem}
\label{t:biasvar1} Let $\pi^0, \pi^1 \in \pz$ have full
supports over $\zstate$.
\begin{romannum}
\item Suppose that the observation sequence $\bfmZ$ is i.i.d.\,
    with marginal $\pi^0$. Then the normalized Hoeffding test
    statistic sequence $\{ n D(\Gamma^n     \| \pi^0) : n\ge
    1\}$ has the following asymptotic bias and variance:
\begin{eqnarray}
\lim_{n \to \infty} \Expect[n D(\Gamma^n \| \pi^0)] &=& \half(N-1)
\label{eqn:bias1}
\\
\lim_{n \to \infty} \Var [n D(\Gamma^n \| \pi^0)]&=&\half(N-1)
\label{eqn:var1}
\end{eqnarray}
where $N = |\zstate|$ denotes the size (cardinality) of
$\zstate$. Furthermore, the following weak convergence result
holds:
\begin{equation}
n D(\Gamma^n \| \pi^0) \xrightarrow[n \to \infty]{d.} \half \chi^2_{N-1} \label{eqn:weak}
\end{equation}
where the right hand side denotes the chi-squared distribution with $N-1$ degrees of freedom.

\item Suppose the sequence $\bfmZ$ is drawn i.i.d. under
    $\pi^1\neq \pi^0$. We then have,
\begin{eqnarray}
\lim_{n \to \infty} \Expect\bigl[n \bigl(D(\Gamma^n \| \pi^0) -
D(\pi^1\| \pi^0) \bigr)\bigr] = \half (N-1)\nonumber
\end{eqnarray}\qed
\end{romannum}
\end{theorem}

The bias result of (\ref{eqn:bias1}) follows from the
unpublished report \cite{clabar89} (see \cite[Sec
III.C]{clabar90}), and the weak convergence result of
(\ref{eqn:weak}) is given in \cite{wil38}. All the results of
the theorem, including (\ref{eqn:var1}) also follow from
\Theorem{ChiSqMM} --- We elaborate on this in section \ref{s:var}.

We see from \Theorem{biasvar1} that the bias of the divergence
statistic $D(\Gamma^n \| \pi^0)$ decays as $\frac{N-1}{2n}$,
irrespective of  whether the observations are drawn from
distribution $\pi^0$ or $\pi^1$.  One could
argue that the problem of high bias in the Hoeffding test
statistic can be addressed by setting a higher threshold.
However, we also notice that when the observations are drawn
under $\pi^0$, the variance of the divergence statistic decays
as $\frac{N-1}{2n^2}$, which can be significant when $N$ is of
the order of $n^2$. This is a more serious flaw of the
Hoeffding test for large alphabet sizes, since it cannot be
addressed as easily.


The weak convergence result in \eqref{eqn:weak}, and other such
results established later in this paper,  can be used to guide the choice of
thresholds for a finite sample test, subject to a constraint on
the probability of false alarm (see for example, \cite[p.
457]{csishi04}).   As an application of \eqref{eqn:bias1}  we
propose the following approximation for the false alarm
probability in the Hoeffding test defined in
\eqref{eqn:Hoefftestnew2},
\begin{equation}
p_\FA\eqdef
\Prob_{\pi^0} \bigl\{ \phi_n^\tH 
		  = 1 \bigr\} \approx  \Prob\bigl\{ \half \sum_{i = 1}^{N-1} W_i^2 \ge n\eta\bigr\}
\label{e:ChiBdd}
\end{equation}
where $\{W_i\}$ are i.i.d.\ $N(0,1)$ random variables.  In this
way we can obtain a simple formula for the threshold to
approximately achieve a given constraint on $p_\FA$. For
moderate values of the sequence length $n$, the $\chi^2$
approximation gives a more accurate prediction of the false
alarm probabilities for the Hoeffding test compared to those
predicted using Sanov's theorem as we demonstrate below.

\begin{figure}[h]
\Ebox{.7}{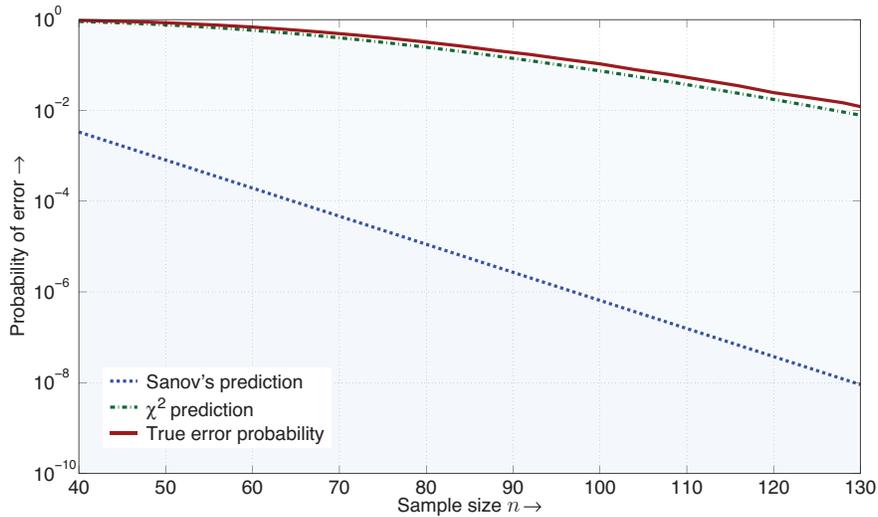}
\caption{\textit{Approximations for the error probability in universal hypothesis testing.}
The error probability of the Hoeffding test is closely approximated by the approximation \eqref{e:ChiBdd}.}
\label{f:errLovely}
\end{figure}

Consider the application of \eqref{e:ChiBdd} in the following
example. We used Monte-Carlo simulations to approximate the
performance of the Hoeffding test described in
(\ref{eqn:Hoefftestnew2}), with  $\pi^0$ the uniform
distribution on an alphabet of size $20$.  Shown in
\Figure{errLovely} is a semi-log plot comparing three
quantities: The probability of false alarm $p_\FA$, estimated
via simulation; the approximation \eqref{e:ChiBdd} obtained
from the Central Limit Theorem; and the approximation obtained
from Sanov's Theorem,  $\log(p_\FA)\approx -n\eta $. It is
clearly seen that the approximation based on the weak
convergence result of \eqref{e:ChiBdd} is \textit{far more
accurate} than the approximation based on Sanov's theorem. It
should be noted that the approximate formula for the false
alarm probability obtained from Sanov's theorem can be made
more accurate by using refinements of large deviation results
given in \cite{ilt95}. However, these refinements are often
difficult to compute. For instance, it can be shown using the
results of \cite{ilt95} that $p_\FA \approx
cn^{\frac{N-3}{2}}\exp(-n\eta)$ where constant $c$ is given by
a surface integral over the surface of the divergence ball,
$\clQ_\eta(\pi^0)$.

One approach to addressing the implementation issues of the
universal test is through clustering (or partitioning)  the
alphabet as in \cite{harremoes09},  or smoothing  in the space
of probability measures as in \cite{zeigut91a,zeigut91b} to
extend the Hoeffding test to the case of continuous alphabets.
The mismatched test proposed here is a generalization of a
partition in the following sense. Suppose that $\{A_i: 1\le
i\le N_a\}$ are disjoint sets satisfying $\cup A_i =\state$,
and let $Y(t)=i$ if $X(t)\in A_i$. Applying \eqref{eqn:var1},
we conclude that the Hoeffding test using $\bfmY$ instead of
$\bfmX$ will have asymptotic variance equal to $\half(N_a-1)$,
where $N_a< N$ for a non-trivial partition. We have:

\begin{proposition}
\label{t:MMPartition} Suppose that the mismatched divergence is
defined with respect to the linear function class
\eqref{eqn:linearfly} using $\psi_i=\ind_{A_i}$,  $1\le i\le
N_a$.   In this case the mismatched test \eqref{eqn:MMtestnew}
coincides with the Hoeffding test using observations $\bfmY$.
\qed
\end{proposition}

The advantage of the mismatched test \eqref{eqn:MMtestnew} over
a partition is that we can incorporate prior knowledge
regarding alternate statistics, and we can include non-standard
`priors' such as continuity of the log-likelihood ratio
function between the null and alternate distributions. This is
useful in anomaly detection applications where one may have
models of anomalous behavior which can be used to design the
correct mismatched test for the desired application.

\subsubsection{Source coding with training}

Let $\pi$ denote a source distribution on a finite alphabet
$\zstate$. Suppose we do not know $\pi$ exactly and we design
optimal codelengths assuming that the distribution is $\mu$:
For letter $z \in \zstate$ we let $\ell(z) = - \log (\mu(z))$
denote Shannon's codeword length. The  expected codelength is
thus,
\[
\Expect[\ell] = \sum_{z \in \zstate} \ell(z) \pi(z) =  H(\pi) + D(\pi \| \mu)
\]
where $H$ denotes the entropy, $-\sum_{z \in\zstate} \pi(z)
\log(\pi(z))$. Let $\ell^* := H(\pi)$ denote the optimal
(minimal) expected codelength.

Now suppose it is known that under $\pi$ the probability of
each letter $z \in \zstate$ is bounded away from zero. That is,
we assume that for some $\epsilon > 0$,
\[
\pi \in \bP_\epsilon := \{\mu \in \pz: \mu(z) > \epsilon, \mbox{ for all } z \in \zstate\}.
\]
Further suppose that a training sequence of length $n$ is
given, drawn under $\pi$. We are interested in constructing a
source code for encoding symbols from the source $\pi$ based on
these training symbols. Let $\Gamma^n$ denote the empirical distribution (i.e., the \textit{type}) of the
observations based on these $n$ training symbols. We assign
codeword lengths to each symbol $z$ according to the following
rule,
\begin{eqnarray*}
\ell(z) = \left\{ \begin{tabular}{cc}$\log \frac{1}{\Gamma^n(z)}$ & if $\Gamma^n \in \bP_{\epsilon/2}$\\
$\log \frac{1}{\pi^\uni(z)}$ & else
\end{tabular} \right.
\end{eqnarray*}
where $\pi^\uni$ is the uniform distribution on $\zstate$.

Let $\clT$ denote the sigma-algebra generated by the training symbols.
The conditional expected codelength given $\clT$
satisfies,
\begin{eqnarray*}
\Expect[\ell^n | \clT] =\left\{ \begin{tabular}{cc}$\ell^* + D(\pi \| \Gamma^n)$ & if $\Gamma^n \in \bP_{\epsilon/2}$\\
$\ell^* + D(\pi \| \pi^\uni)$& else
\end{tabular} \right.
\end{eqnarray*}
We study the behavior of $\Expect[\ell^n- \ell^* | \clT]$ as a
function of $n$. We argue in the appendix that a modification
of the results from Theorem~\ref{t:ChiSqMM} can be used to
establish the following relations:
\begin{eqnarray}\label{eqn:srccoding}
n (\Expect[\ell^n | \clT] -  \ell^*) &\xrightarrow[n \to \infty]{d.}& \half \chi^2_{N-1}\nonumber \\
\Expect[n(\ell^n  -  \ell^*)] &\xrightarrow[n \to \infty]{}& \half(N-1) \\
\Var[n \Expect[\ell^n | \clT] ] &\xrightarrow[n \to \infty]{}& \half (N-1) \nonumber
\end{eqnarray}
where $N$ is the cardinality of the alphabet $\zstate$.
Comparing with \Theorem{biasvar1} we conclude that the
asymptotic behavior of the excess codelength is identical to
the asymptotic behavior of the Hoeffding test statistic
$D(\Gamma^n \| \pi)$ under $\pi$. Methods such as those
proposed in this paper can be used to reduce high variance,
just as in the hypothesis testing problem emphasized in this
paper.

\subsubsection{Filtering}


\def\haB{\widehat{B}}

The recent paper \cite{zhofumar10} considers approximations for the nonlinear filtering problem.   Suppose that $\bfmX$ is a Markov chain on $\Re^n$,   and $\bfmY$ is an associated observation process on $\Re^p$ of the form $Y(t)=\gamma(X(t),W(t))$, where $\bfmW$ is an i.i.d.\ sequence.   The conditional distribution of $X(t)$ given $\{Y(0),\dots,Y(t)\}$ is denoted $B_t$; known as the \textit{belief state} in this literature.  The evolution of the belief state can expressed in a recursive form:  For some mapping $\phi\colon\clB(\Re^n)\times \Re^p\to \clB(\Re^n)$,
\[
B_{t+1} = \phi(B_t,Y_{t+1}),\qquad t\ge 0
\]

The approximation proposed in  \cite{zhofumar10} is based
aprojection of $B_t$ onto an exponential family of densities
over $\Re^n$,  of the form $p_\theta(x) = p_0(x)
\exp({\theta^\transpose \psi(x) -\Lambda(\theta)})$,
$\theta\in\Re^d$. They consider the \textit{reverse
$I$-projection},
\[
B^\varrho =\argmin_{\mu\in\clE} D( B\| \mu)
\]
where the minimum is over $\clE =\{   p_\theta \}$.  From the
definition of divergence this is equivalently expressed,
\begin{equation}
B^\varrho =\argmax_\theta  \int\Bigl( \theta^\transpose \psi(x) -\Lambda(\theta) \Bigr) \, B(dx)
\label{e:Brho}
\end{equation}
A projected filter is defined by the recursion,
\begin{equation}
\haB_{t+1} = [\phi(\haB_t,Y_{t+1}) ]^\varrho,\qquad t\ge 0
\label{e:projFilter}
\end{equation}
The techniques in the current paper provide algorithms for
computation of this projection, and suggest alternative
projection schemes, such as the robust approach described in
\Section{s:log-linear}.

\subsection{Basic structure of mismatched divergence}

The mismatched test is defined to be a relaxation of the
Hoeffding test described in (\ref{eqn:Hoefftestnew2}). We
replace the divergence functional with the mismatched
divergence $\DMM(\Gamma^n \| \pi^0)$ to obtain the mismatched test
sequence,
\begin{equation}
\begin{aligned}
\phi_n^\MM
&= \ind\{\DMM(\Gamma^n \| \pi^0) \geq \eta \}
= \ind\{\Gamma^n \notin \QMM_\eta(\pi^0) \}
\label{eqn:MMtest}
\end{aligned}
\end{equation}
where $\QMM_\eta(\pi^0)$ is the mismatched divergence ball of
radius $\eta$ around $\pi^0$ defined analogously to
(\ref{e:Qball}):
\begin{equation}
\QMM_\eta(\mu) = \{\nu \in \pz: \DMM(\nu \| \mu) < \eta\}. \label{e:QMMball}
\end{equation}

The next proposition establishes some basic geometry of the mismatched divergence balls.
For any function $g$ we define the following hyperplane and half-space:
\begin{equation}
\begin{aligned}
\clH_g& \eqdef
            \{ \nu : \nu(g)=0\}
\\
\clH^-_g& \eqdef
            \{ \nu : \nu(g)<0\}.
\end{aligned}
\label{e:clH}
\end{equation}

\begin{proposition}
\label{t:GeoMM}
The following hold for any $\nu,\pi\in\pz$, and any  collection of functions $\clF$:
\begin{romannum}
\item  For each $\eta>0$ we have $\QMM_\eta(\pi) \subset
    \bigcap\clH^-_g $, where the intersection is over all
    functions $g$ of the form,
\begin{equation}
g=f - \Lambda_{\pi}(f) -\eta
\label{e:g}
\end{equation}
with $f\in\clF$.

\item Suppose that $\eta=  \DMM(\nu\| \pi)$ is finite and
    non-zero. Further suppose that  for $\nu^1
        = \nu$ and $\nu^2 = \pi$, the supremum in
    \eqref{eqn:DivMMclF} is achieved by $f^*\in\clF$.
    Then $\clH_{g^*} $ is a supporting hyperplane to $ \QMM_\eta(\pi)
    $, where $g^*$ is given in \eqref{e:g} with $f=f^*$.
\end{romannum}
\end{proposition}

\begin{IEEEproof}
(i)  Suppose  $\mu\in\QMM_\eta(\pi)$. Then, for any $f\in\clF$,
\[
\mu(f) - \Lambda_{\pi}(f) -\eta \le \DMM(\mu \| \pi) -\eta <0
\]
That is, for any  $f\in\clF$, on defining $g$ by \eqref{e:g} we obtain the desired inclusion $\QMM_\eta(\pi) \subset \clH_{g}^-$.

(ii) Let $\mu \in \clH_{g^*}$ be arbitrary. Then we have:
\begin{eqnarray*}
\DMM(\mu \| \pi) &=& \sup_{r}  \bigl(\mu(f_r) - \Lambda_{\pi}(f_r) \bigr)\\
&\geq& \mu(f^*) - \Lambda_{\pi}(f^*) \\&=& \Lambda_\pi(f^*) + \eta - \Lambda_\pi(f^*) =  \eta.
\end{eqnarray*}
Hence it follows that $\clH_{g^*}$ supports $\QMM_\eta(\pi)$ at
$\nu$.

\end{IEEEproof}

\begin{figure}[h]
\Ebox{.5}{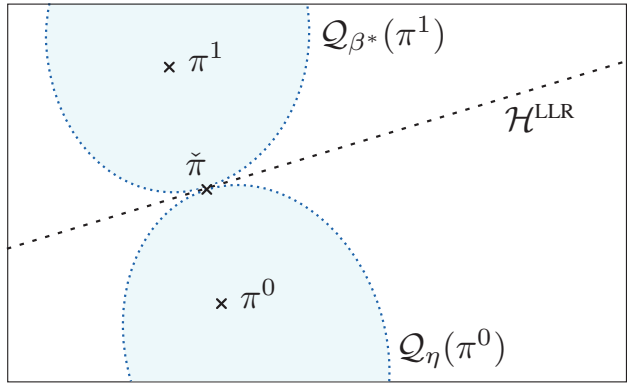}
\caption{\textit{Geometric interpretation of the log likelihood ratio test.}
The exponent $ \beta^*=\beta^*(\eta)$ is the largest constant satisfying $\clQ_\eta(\pi^0)\cap\clQ_{\beta^*}(\pi^1)=\emptyset$.  The  hyperplane $\HLLR \eqdef \{\nu: \nu(L) = \cpi(L)\}$
separates the convex sets $\clQ_\eta(\pi^0)$ and
$\clQ_{\beta^*}(\pi^1)$.}
\label{f:geometryLRT}
\end{figure}

\subsection{Asymptotic optimality of the mismatched test}
The asymptotic performance of a binary hypothesis testing
problem is typically characterized in terms of error exponents.
We adopt the following criterion for performance evaluation,
following Hoeffding~\cite{hoe65a} (and others, notably
\cite{zeigut91a,zeigut91b}.) Suppose that the observations
$\bfmZ = \{Z_t : t=1,\ldots \}$ form an i.i.d.\ sequence
evolving on $\zstate$. For a given $\pi^0$, and a given
alternate distribution $\pi^1$, the type I and type II error
exponents are denoted respectively by,
\begin{equation}
\begin{aligned}
J^0_{\phi}  & \eqdef \liminf_{n \to \infty} -\frac{1}{n}\log(\Prob_{\pi^0}\{
\phi_n=1\} ),
\breakMed
J^1_{\phi}   & \eqdef    \liminf_{n \to \infty} -\frac{1}{n}\log(\Prob_{\pi^1}\{
\phi_n=0\} )
\end{aligned}
\label{eqn:ErrorProb}
\end{equation}
where in the first limit the marginal distribution of $Z_t$ is
$\pi^0$, and in the second it is $\pi^1$. The limit $J^0_{\phi}
$ is also called the false-alarm error exponent,  and
$J^1_{\phi} $ the missed-detection error exponent.

For a given constraint $\eta > 0$ on the false-alarm exponent
$J^0_{\phi}$, an optimal test is the solution to the asymptotic
Neyman-Pearson hypothesis testing problem,
\begin{equation}
\beta^*(\eta)=   \sup \{   J^1_{\phi}   :        \ \mbox{\it subject to} \  J^0_{\phi}
\ge \eta\}  \label{eqn:GameP1}
\end{equation}
where the supremum is over all allowed test sequences $\bfphi$.
While the exponent $\beta^*(\eta)=\beta^*(\eta,\pi^1)$ depends
upon $\pi^1$, Hoeffding's test we described in
(\ref{eqn:Hoefftestnew2}) does not require knowledge of
$\pi^1$, yet achieves the optimal exponent
$\beta^*(\eta,\pi^1)$ for any $\pi^1$. The optimality of
Hoeffding's test established in \cite{hoe65a} easily follows
from Sanov's theorem.

While the mismatched test described in (\ref{eqn:MMtestnew2})
is not always optimal for (\ref{eqn:GameP1}) for a general
choice of $\pi^1$, it is optimal for some specific choices of
the alternate distributions. The following corollary to
Proposition~\ref{t:GeoMM} captures this idea.

\begin{corollary}
\label{t:MMoptwithLLR} Suppose $\pi^0, \pi^1 \in \pz$ have
equal supports. Further suppose that for all $\alpha > 0$,
there exists $\tau \in \Re$ and $r \in \Re^d$ such that
\[
\alpha L(z) + \tau = f_r(z) \quad \mbox{a.e.} \quad [\pi^0],
\]
where $L$ is the log likelihood-ratio function $L \eqdef
\log(\pi^1/\pi^0)$. Then the mismatched test is optimal in the
sense that the constraint $J^0_{\phi^{\hbox{\rm\tiny MM}}} \ge
\eta$ is satisfied with equality,  and under $\pi^1$ the
optimal error exponent is achieved; i.e.
$J^1_{\phi^{\hbox{\rm\tiny MM}}} = \beta^*(\eta)$ for all $\eta
\in (0, D(\pi^1 \| \pi^0))$.
\end{corollary}

\begin{IEEEproof}
Suppose that the conditions stated in the corollary hold.
Consider the twisted distribution $\cpi = \kappa
(\pi^0)^{1-\varrho} (\pi^1)^\varrho$, where $\kappa$ is a
normalizing constant and $\varrho \in (0,1)$ is chosen so as to
guarantee $D(\cpi \|\pi^0) = \eta$. It is known that  the
hyperplane $\HLLR \eqdef \{\nu: \nu(L) = \cpi(L)\}$ separates
the divergence balls $\clQ_\eta(\pi^0)$ and
$\clQ_{\beta^*}(\pi^1)$ at $\cpi$. This geometry, which is
implicit in~\cite{zeigut91a}, is illustrated in
\Figure{geometryLRT}.

From the form of $\cpi$ it is also
clear that
\[
\log \frac{\cpi}{\pi^0} = \varrho L - \Lambda_{\pi^0} (\varrho L).
\]
Hence it follows that the supremum in the variational
representation of $D(\cpi \| \pi^0)$ is achieved by $\varrho
L$. Furthermore, since $\varrho L + \tau \in \clF$ for some
$\tau \in \Re$ we have
\begin{eqnarray*}
\DMM(\cpi \| \pi^0) &=& D(\cpi \| \pi^0) = \eta \\
&=& \cpi(\varrho L + \tau) - \Lambda_{\pi^0}(\varrho L + \tau)\\ &=& \cpi(\varrho L) - \Lambda_{\pi^0}(\varrho L).
\end{eqnarray*}
This means that $\HLLR = \{\nu: \nu(\varrho L - \Lambda_{\pi^0}
(\varrho L) -\eta  ) = 0\}$. Hence, by applying Proposition
\ref{t:GeoMM} (ii) it follows that the hyperplane $\HLLR$
separates $\QMM_\eta(\pi^0)$ and $\clQ_{\beta^*}(\pi^1)$. This
in particular means that the sets $\QMM_\eta(\pi^0)$ and
$\clQ_{\beta^*}(\pi^1)$ are disjoint. This fact, together with
Sanov's theorem proves the corollary.
\end{IEEEproof}
The corollary indicates that while using the mismatched test in
practice, the function class might be chosen to include
approximations to scaled versions of the log-likelihood ratio
{functions} of the anticipated alternate distributions
$\{\pi^1\}$ with respect to $\pi^0$.

The mismatched divergence has several equivalent
characterizations.  We first relate it to an ML estimate from a
parametric family of distributions.

\subsection{Mismatched divergence and ML estimation}
On interpreting $f_r - \Lambda_\pi(f_r) $ as a log-likelihood
ratio {function} we obtain in \Proposition{Pythag} the
following representation of mismatched divergence,
\begin{equation}
\DMM(\mu \| \pi)=  \sup_{r \in \Re^d}  \bigl(\mu(f_r) - \Lambda_{\pi}(f_r) \bigr) = D (\mu \| \pi ) - \inf_{\nu \in \clE_\pi}D(\mu \| \nu).
\label{e:MMreverse}
\end{equation}
The infimum on the RHS of \eqref{e:MMreverse} is known as \textit{reverse I-projection} \cite{csishi04}.
\Proposition{MM=ML} that follows uses this representation to obtain other interpretations of the mismatched test.

\begin{proposition}
\label{t:Pythag}
The identity \eqref{e:MMreverse} holds for any function class $\clF$.
The supremum is achieved by some $r^* \in \Re^d$ if and only if
the infimum is attained at $\nu^*= \cpi^{r^*}\in  \clE_\pi$. If a minimizer
$\nu^*$ exists, we obtain the {\em generalized Pythagorean identity},
\[
 D (\mu \| \pi ) =
\DMM(\mu \| \pi) +D(\mu \| \nu^*)
\]
\end{proposition}

\begin{IEEEproof}
For any $r$ we have $\mu(f_r) - \Lambda_\pi(f_r) =
\mu(\log(\cpi^r/\pi))$.  Consequently,
\[
\begin{aligned}
\DMM(\mu \| \pi) &= \sup_{r} \bigl(\mu(f_r) - \Lambda_\pi(f_r) \bigr)
\\
&= \sup_{r} \mu \left( \log\left(\frac{\mu}{\pi}\frac{\cpi^r}{\mu} \right)\right)
\\
&=  \sup_{r}   \left\{  D (\mu \| \pi ) -  D(\mu \| \cpi^r) \right\}
\end{aligned}
\]
This proves the identity \eqref{e:MMreverse}, and the remaining conclusions follow directly.
\end{IEEEproof}

The representation of \Proposition{Pythag} invites the
interpretation of the optimizer in the definition of the
mismatched test statistic in terms of an ML estimate.  Given
the well-known correspondence between maximum-likelihood
estimation and the  generalized likelihood ratio test (GLRT),
\Proposition{MM=ML} implies that the mismatched test is a
special case of the GLRT analyzed in \cite{zeizivmer92}.

\begin{proposition}
\label{t:MM=ML} Suppose that the observations $\bfmZ$ are
modeled as an i.i.d.\ sequence,  with marginal in the family
$\clE_\pi$. Let $\har^n$ denote the ML estimate of $r$ based on
the first $n$ samples,
\begin{eqnarray*}
\har^n&\in& \argmax_{r\in\Re^d} \Prob_{\cpi^r}\{ Z_1 =a_1, Z_2=a_2, \ldots, Z_n=a_n \}\\
&&=
 \argmax_{r\in\Re^d} \displaystyle \Pi_{i=1}^n \cpi^r(a_i)
\end{eqnarray*}

where $a_i$ indicates the observed value of the $i$-th symbol.
Assuming the maximum is attained we have the following interpretations:
\begin{romannum}
\item
The distribution $ \cpi^{\har^n}$ solves the reverse I-projection problem,
\[
\cpi^{\har^n} \in \argmin_{\nu \in \clE_\pi} D(\Gamma^n \| \nu).
\]
\item
The function $f^*=f_{\har^n}$
achieves the supremum that defines the mismatched divergence,
$\DMM(\Gamma^n\| \pi) =\Gamma^n(f^*) - \Lambda_{\pi}(f^*)$.

\end{romannum}
\end{proposition}

\begin{IEEEproof}
The ML estimate can be expressed $\har^n =\argmax_{r\in\Re^d}
\langle \Gamma^n, \log \cpi^r \rangle$, and hence (i) follows
by the identity,
\[
 \argmin_{\nu \in \clE_\pi} D(\Gamma^n \| \nu) = \argmax_{\nu \in \clE_\pi} \langle \Gamma^n, \log \nu \rangle,\qquad \nu\in\clP.
\]
Combining the result of part (i) with \Proposition{Pythag} we
get the result of part (ii).
\end{IEEEproof}
From conclusions of \Proposition{Pythag} and
\Proposition{MM=ML} we have,
\begin{eqnarray*}
\DMM(\Gamma^n \| \pi ) &=& \langle \Gamma^n,\log \frac{\cpi^{\har^n}}{\pi} \rangle \\
&=& \max_{\nu \in \clE_{\pi}} \langle \Gamma^n,\log \frac{\nu}{\pi} \rangle\\
&=& \max_{\nu \in \clE_{\pi}} \frac{1}{n} \sum_{i=1}^n \log \frac{\nu(Z_i)}{\pi(Z_i)}.
\end{eqnarray*}
In general when the supremum in the definition of
$\DMM(\Gamma^n \| \pi)$ may not be achieved, the maxima in the
above equations are replaced with suprema and we have the
following identity:
\[
\DMM(\Gamma^n \| \pi ) = \sup_{\nu \in \clE_{\pi}} \frac{1}{n} \sum_{i=1}^n \log \frac{\nu(Z_i)}{\pi(Z_i)}.
\]
Thus the test statistic used in the mismatched test of
(\ref{eqn:MMtestnew2}) is exactly the generalized likelihood
ratio between the family of distributions $\clE_{\pi^0}$ and
$\pi^0$ where
\[
\clE_{\pi^0} = \{\pi^0 \exp(f_r - \Lambda_{\pi^0}(f_r)): r \in \Re^d \}.
\]

More structure can be established when the function class is linear.

\subsection{Linear function class and I-projection}

The mismatched divergence   introduced in
\cite{abbmedmeyzhe07a} was restricted to a linear function
class. Let $\{\psi_i:1\le i\le d\}$ denote $d$ functions on
$\zstate$,
let $\psi=(\psi_1,\dots,\psi_d)^\transpose$,
and let
$f_r= r^\transpose \psi$ in the definition \eqref{e:clF}:
\begin{equation}
\clF = \Bigl\{f_r = \sum_{i=1}^d r_i \psi_i: r \in \Re^d \Bigr\}.
\label{eqn:linearfly}
\end{equation}
A linear function class is particularly appealing because the
optimization problem in (\ref{eqn:DivMMclF}) used to define the
mismatched divergence becomes a convex program and hence is
easy to evaluate in practice. Furthermore, for such a linear
function class, the collection of twisted distributions
$\clE_\pi$ defined in (\ref{eqn:family}) forms an
\textit{exponential family} of distributions.

\Proposition{Pythag} expresses  $\DMM(\mu \| \pi)$ as a
difference between the ordinary divergence and the value of a
reverse I-projection $ \inf_{\nu \in \clE_\pi}D(\mu \| \nu) $.
The next result establishes a characterization in terms of a
(forward) I-projection.  For a given vector $c\in\Re^d$ we let
$\bP$ denote  the \textit{moment class}
\begin{equation}
\bP=\{\nu \in \pz: \nu(\psi)=c\}
\label{e:bP}
\end{equation}
where $\nu(\psi) = (\nu(\psi_1), \nu(\psi_2), \ldots,
\nu(\psi_d))^\transpose$.

\begin{proposition}
\label{t:revIprojPyth}
Suppose that the supremum in the definition of $\DMM(\mu \| \pi)$ is
achieved at some  $r^*\in\Re^d$.   Then,
\begin{romannum}

\item
The distribution $\nu^*\eqdef\cpi^{r^*} \in\clE_\pi$ satisfies,
\[
 \DMM(\mu \| \pi) =
D(\nu^*\| \pi)  = \min\{ D(\nu \| \pi) : \nu\in\bP \},
\]
where $\bP$ is defined using $c=  \mu(\psi)$ in (\ref{e:bP}).

\item
$\displaystyle
 \DMM(\mu \| \pi) = \min\{ D(\nu \| \pi) : \nu\in \clH_{g^*}\}$,   where $g^*$ is given in \eqref{e:g} with $f={r^*}^\transpose \psi$,  and $\eta =  \DMM(\mu \| \pi) $.

\end{romannum}
\end{proposition}
\begin{IEEEproof}
Since the supremum is achieved, the gradient must vanish by
the first order condition for optimality:
\[
\nabla \bigl(\mu(f_r) - \Lambda_{\pi}(f_r) \bigr)\Big|_{r=r^*} =0
\]
The gradient is computable, and the identity above can thus be
expressed  $\mu(\psi) - \cpi^{r^*}(\psi)=0$.  That is,  the
first order condition for optimality is equivalent to the
constraint $\cpi^{r^*}\in\bP$. Consequently,
\begin{eqnarray*}
D(\nu^* \| \pi) &=& \langle \cpi^{r^*}, \log \frac{\cpi^{r^*}}{\pi} \rangle \\
&=& \cpi^{r^*} ({r^*}^{\transpose} \psi) - \Lambda_\pi({r^*}^{\transpose} \psi) \\
&=& \mu ({r^*}^{\transpose} \psi) - \Lambda_\pi({r^*}^{\transpose} \psi)= \DMM(\mu \| \pi).
\end{eqnarray*}
Furthermore, by the convexity of $\Lambda_\pi(f_r)$ in $r$, it
follows that the optimal $r^*$ in the definition of $\DMM(\nu
\| \pi)$ is the same for all $\nu \in \bP$. Hence, it follows
by the Pythagorean equality of Proposition \ref{t:Pythag} that
\[
D(\nu \| \pi) = D(\nu \| \nu^*) + D(\nu^* \| \pi), \mbox{ for all } \nu \in \bP.
\]
Minimizing over $\nu \in \bP$ it follows that $\nu^*$ is the
I-projection of $\pi$ onto $\bP$:
\[
D(\nu^*\| \pi)  = \min\{ D(\nu \| \pi) : \nu\in\bP \}
\]
which gives (i).

To establish (ii), note first that by (i) and the inclusion  $\bP\subset\clH_{g^*} $ we have,
\begin{eqnarray*}
 \DMM(\mu \| \pi) &=&  \min\{ D(\nu \| \pi) : \nu\in\bP \}\\ &\ge&  \inf\{ D(\nu \| \pi) : \nu\in \clH_{g^*} \}.
\end{eqnarray*}
The reverse inequality follows from    \Proposition{GeoMM}~(i), and moreover the infimum is achieved with $\nu^*$.
\end{IEEEproof}

\begin{figure}[h]
\Ebox{.5}{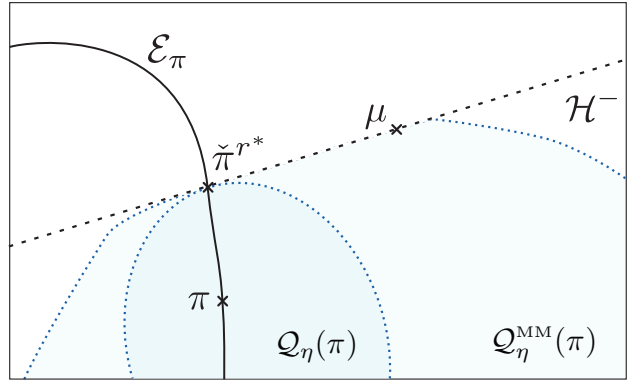}
\caption{\textit{Interpretations of the mismatched divergence for a linear function class.}
The distribution  $  \cpi^{r^*}$  is the I-projection of $\pi$ onto a hyperplane $\clH_{g^*}$.   It is also the reverse I-projection of $\mu$ onto the exponential family $ \clE_\pi$.      }
\label{f:geometryMM_H+}
\end{figure}

The geometry underlying mismatched divergence for a linear function class is illustrated in \Figure{geometryMM_H+}.
Suppose that the assumptions of  \Proposition{revIprojPyth}  hold, so that the supremum in
\eqref{e:MMreverse} is achieved at $r^*$. Let  $\eta =\DMM(\mu \| \pi)=   \mu(f_{r^*}) - \Lambda_{\pi}(f_{r^*}) $,   and $g^*=f_{r^*}-\bigl(\eta
+\Lambda_{\pi}(f_{r^*}) \bigr)$. \Proposition{GeoMM} implies that $\clH_{g^*}$ defines a
hyperplane passing through $\mu$, with $\clQ_\eta(\pi)\subset\QMM_\eta(\pi)\subset\clH^-_{g^*}$.
This is strengthened in the linear case by
\Proposition{revIprojPyth}, which states that  $\clH_{g^*}$
supports $\clQ_\eta(\pi)$ at  the   distribution $  \cpi^{r^*}$.  Furthermore \Proposition{Pythag} asserts that the
distribution $  \cpi^{r^*}$  minimizes $D(\mu\|\cpi)$ over all $\cpi\in \clE_\pi$.

The result established in Corollary \ref{t:MMoptwithLLR} along
with the interpretation of the mismatched test as a GLRT can be
used to show that the GLRT is asymptotically optimal for an
exponential family of distributions.
\begin{theorem}
\label{t:GLRToptimal} Let $\pi^0$ be some probability
distribution over a finite set $\zstate$. Let $\clF$ be a
linear function class as defined in (\ref{eqn:linearfly}) and
$\clE_{\pi^0}$ be the associated exponential family of
distributions defined in (\ref{eqn:family}). Consider the
generalized likelihood ratio test (GLRT) between $\pi^0$ and
$\clE_{\pi^0}$ defined by the following sequence of decision
rules:
\[
\phi_n^\tG = \ind\{\sup_{\nu \in \clE_{\pi^0}} \frac{1}{n} \sum_{i=1}^n \log \frac{\nu(Z_i)}{\pi^0(Z_i)} \geq \eta \}.
\]
The GLRT solves the composite hypothesis testing problem
(\ref{eqn:GameP1}) for all $\pi^1 \in \clE_{\pi^0}$ in the
sense that the constraint $J^0_{\phi^\tG} \ge \eta$ is
satisfied with equality, and under $\pi^1$ the optimal error
exponent $\beta^*(\eta)$ is achieved for all $\eta \in (0,
D(\pi^1 \| \pi^0))$ and for all $\pi^1 \in \clE_{\pi^0}$; i.e.,
$J^1_{\phi^\tG} = \beta^*(\eta)$.
\end{theorem}
\begin{IEEEproof}
From Proposition \ref{t:MM=ML} and the discussion following the
proposition, we know that $\phi^\tG$ is the same as the
mismatched test defined with respect to the function class
$\clF$. Moreover, any distribution $\pi^1 \in \clE_{\pi^0}$  is
of the form $\cpi^r = \pi^0 \exp(f_r - \Lambda_{\pi^0}(f_r) )$
for some $r \in \Re^d$ as defined in (\ref{eqn:twisteddistrn}).
Using $L$ to denote the log-likelihood ratio {function} between
$\pi^1$ and $\pi^0$, it follows by the linearity of $\clF$ that
for any $\alpha> 0$,
\begin{eqnarray*}
\alpha L &=& \alpha (f_r - \Lambda_{\pi^0}(f_r) )\\
&=& f_{\alpha r} + \tau
\end{eqnarray*}
for some $\tau \in \Re$. Hence, it follows by the conclusion of
Corollary \ref{t:MMoptwithLLR} that the GLRT $\phi^\tG$ solves
the composite hypothesis testing problem (\ref{eqn:GameP1})
between $\pi^0$ and $\clE_{\pi^0}$.
\end{IEEEproof}
The above result is a special case of the sufficient conditions
for optimality of the GLRT established in \cite[Thm 2, p.
1600]{zeizivmer92}. From the proof it is easily seen that the
result can be extended to hold for composite hypothesis tests
between $\pi^0$ and any family of distributions $\clE_{\pi^0}$
of the form in (\ref{eqn:family}) provided $\clF$ is closed
under positive scaling. It is also possible to strengthen the
result of \Corollary{MMoptwithLLR} to obtain an alternate proof
of \cite[Thm 2, p. 1600]{zeizivmer92}.
We refer the reader to \cite{unn10} for details.

\subsection{Log-linear function class and robust hypothesis testing}
\label{s:log-linear}

In the prior work \cite{panmeyvee04a,panmey06a} the following
relaxation of entropy is considered,
\begin{equation}
\DR(\mu \| \pi) \eqdef \inf_{\nu \in \bP} D(\mu \| \nu)
 \label{e:supMM}
\end{equation}
where the moment class $\bP$ is defined in \eqref{e:bP} with
$c=\pi(\psi)$, for a given  collection of functions
$\{\psi_i : 1\leq i \leq d\}$.   The associated universal test
solves a min-max robust hypothesis testing problem.

We show here that $\DR$ coincides with $\DMM$ for a particular function class.  It is described as \eqref{e:clF} in which each function $f_r$ is of the log-linear form,
\[
f_r = \log(1+r^\transpose \psi)
\]
subject to the constraint that $ 1+r^\transpose \psi (z)$ is
strictly positive for each $z$. We further require that the
functions $\psi$ have zero mean under distribution $\pi$ -
i.e., we require $\pi(\psi) = 0$.

\begin{proposition}
\label{t:Drob} For a given $\pi\in\pz$, suppose that the
log-linear function class  $\clF$ is chosen with functions
$\{\psi_i\}$ satisfying $\pi(\psi)=0$.   Suppose that the
moment class used  in the definition of $\DR$ is chosen
consistently, with $c=0$ in (\ref{e:bP}).  We then have for
each $\mu\in\pz$,
\[
\DMM(\mu\| \pi) =
\DR(\mu\| \pi)
\]
\end{proposition}

\begin{IEEEproof}
For each $\mu \in \pz$, we obtain the following identity by applying Theorem 1.4 in \cite{panmey06a},
\begin{eqnarray*}
{\inf_{\nu \in \bP} D(\mu \| \nu)} = \sup \{\mu(\log(1 + r^\transpose \psi)): 1 + r^\transpose \psi (z)>0 \mbox{ for all } z\in\zstate\}
\end{eqnarray*}
Moreover, under the assumption that $\pi(\psi)=0$ we obtain,
\[
\Lambda_{\pi}(\log(1
+ r^\transpose \psi))=\log(\pi(1 + r^\transpose \psi))=0
\]
Combining these identities gives,
\begin{eqnarray*}
{\DR(\mu\|\pi)}
& \eqdef &
\inf_{\nu \in \bP} D(\mu \| \nu)
\nonumber\\
&=&
\sup\left\{\mu(\log(1 + r^\transpose\psi)) -\Lambda_{\pi}(\log(1 + r^\transpose \psi)): \right.\\
&& \left. \qquad 1 + r^\transpose \psi (z)>0 \mbox{ for all } z\in\zstate \right\}
\nonumber\\
&=&
\sup_{f \in \clF} \bigl\{\mu(f)-\Lambda_{\pi}(f) \bigr\}=\DMM(\mu\|\pi)
\nonumber
\end{eqnarray*}
\end{IEEEproof}

\section{Asymptotic Statistics}\label{s:var}

In this section, we analyze the asymptotic statistics of the
mismatched test. We require some assumptions regarding the
function class $\clF = \{f_r: r\in \Re^d\}$  to establish these
results. Note that the second and third assumptions given below
involve a distribution $\mu^0 \in \pz$, and a vector $s \in
\Re^d$. We will make specialized versions of these assumptions
in establishing our results, based on specific values of
$\mu^0$ and $s$. We use $\zstate_{\mu^0} \subset \zstate$ to
denote the support of $\mu^0$ and $\clP(\zstate_{\mu^0})$ to
denote the space of probability measures supported on
$\zstate_{\mu^0}$, viewed as a subset of $\pz$.

\medbreak

\begin{quote}
{\large
\textbf{Assumptions}}%
\nobreak%
\begin{assumption}
\item
\label{ass:frisC2}
 $f_r(z)$  is $C^2$ in $r$ for each $z\in\zstate$.

\item \label{ass:supunique} There exists a neighborhood
    $B$ of $\mu^0$, open in $\clP(\zstate_{\mu^0})$ such that for each $\mu \in B$,
    the supremum in the definition of $\DMM(\mu\|\mu^0)$ in
    \eqref{eqn:DivMMclF} is achieved at a unique point
    $r(\mu)$.

\item
\label{ass:nondegenerate} The vectors $\{\psi_0,\dots,\psi_d\}$ are
linearly independent over the support of $\mu^0$, where
$\psi_0\equiv 1$,  and for each $i\ge 1$
\begin{equation}
\psi_i(z) =\frac{\partial}{\partial r_i} f_r(z) \Big|_{r = s},\qquad z\in\zstate.
\label{e:psipi}
\end{equation}
\end{assumption}
\end{quote}
The linear-independence assumption in \Ass{nondegenerate} is
defined as follows:  If there are constants $\{a_0,\dots,a_d\}$
satisfying $\sum_{i = 1}^d a_i \psi_i(z) = 0$  a.e.\ $[\mu^0]$,
then $a_i=0$ for each $i$. In the case of a linear function
class, the functions $\{\psi_i, i\geq 1\}$ defined in
(\ref{e:psipi}) are just the basis functions in
(\ref{eqn:linearfly}). \Lemma{Siginv} provides an alternate
characterization of Assumption~\Ass{nondegenerate}.

For any $\mu\in\pz$  define the covariance matrix $\Sigma_\mu$ via,
\begin{equation}
\Sigma_\mu(i,j) = \mu(\psi_i\psi_j) - \mu(\psi_i)\mu(\psi_j), \qquad 1 \leq i,j \leq d.
\label{e:Sigmapi}
\end{equation}
We use $\Cov_{\mu}(g)$ to denote the covariance of an arbitrary
real-valued function $g$ under $\mu$:
\begin{equation}
\Cov_{\mu}(g) \eqdef \mu(g^2) - \mu(g)^2.
\label{e:Cov}
\end{equation}

\begin{lemma}
\label{t:Siginv}
Assumption~\Ass{nondegenerate} holds if and only if $\Sigma_{\mu^0}>0$.
\end{lemma}
\begin{IEEEproof}
We evidently have $v^\transpose \Sigma_{\mu^0} v = \Cov_{\mu^0}(v^\transpose \psi)\ge 0$
for any vector $v \in \Re^d$.  Hence, we have the following
equivalence:  For any $v \in \Re^d$, on denoting $c_v=\mu^0(v^\transpose \psi)$,
\[
v^\transpose \Sigma_{\mu^0} v = 0 \quad \Leftrightarrow \quad \sum_{i=1}^d v_i \psi_i(z) = c_v \ \  \mbox{ a.e. } [\mu^0]
\]
The conclusion of the lemma follows.
\end{IEEEproof}

We now present our main asymptotic results. \Theorem{ChiSqMM}
identifies  the asymptotic bias and variance of the mismatched
test statistic under the null hypothesis, and also under the
alternate hypothesis.    A key observation is that the
asymptotic bias and variance does not depend on $N$, the
cardinality of $\zstate$.



\begin{theorem}
\label{t:ChiSqMM} Suppose that the observation sequence $\bfmZ$
is i.i.d.\, with marginal $\pi$. Suppose that there exists
$r^*$ satisfying $f_{r^*} = \log(\pi / \pi^0)$. Further,
suppose that Assumptions \Ass{frisC2}, \Ass{supunique},
\Ass{nondegenerate} hold with $\mu^0 = \pi$ and $s = r^*$.
Then,
\begin{romannum}
\item
When $\pi=\pi^0$,
\begin{eqnarray}
\lim_{n \to \infty} \Expect[n \DMM(\Gamma^n \| \pi^0)] &=& \half d\label{e:part1bias}\\
\lim_{n \to \infty} \Var [n \DMM(\Gamma^n \| \pi^0)]&=&\half d\label{e:part1var}\\
n \DMM(\Gamma^n \| \pi^0) &\xrightarrow[n \to \infty]{d.}& \half
\chi^2_{d} \label{e:part1dis}\nonumber
\end{eqnarray}

\item
 When $\pi=\pi^1\neq \pi^0$,  we
have with $\sigma^2_1 \eqdef \Cov_{\pi^1} (f_{r^*} )$,
\begin{eqnarray}
\lim_{n \to \infty} \Expect[n (\DMM(\Gamma^n \| \pi^0) - D(\pi^1\| \pi^0))] &=& \half d\label{e:part2bias}\\
\lim_{n \to \infty}\Var [n^\half \DMM(\Gamma^n \| \pi^0)]
&=& \sigma^2_1\label{e:part2var}
\end{eqnarray}
\vspace{-0.25in}
\begin{eqnarray}
n^\half (\DMM(\Gamma^n \| \pi^0) - D(\pi^1 \| \pi^0) )
\xrightarrow[n \to \infty]{d.} \clN (0,\sigma^2_1
)\label{e:part2dis}.
\end{eqnarray}
\end{romannum} \qed
\end{theorem}
In part (ii) of \Theorem{ChiSqMM}, the assumption that $r^*$
exists implies that $\pi^1$ and $\pi^0$ have equal supports.
Furthermore, if Assumption~\Ass{nondegenerate} holds in part
(ii), then a sufficient condition for
Assumption~\Ass{supunique} is that the function $V(r) \eqdef
(-\pi^1(f_r) + \Lambda_{\pi^0}(f_r))$ be coercive in $r$.  And,
under \Ass{nondegenerate}, the function $V$ is strictly convex
and coercive   in the following settings:   (i) If the function
class is linear, or (ii)  the function class is log-linear, and
the two distributions $\pi^1$ and $\pi^0$ have common support.
We use this fact in \Theorem{Linearoffset} for the linear
function class. The assumption of the existence of $r^*$
satisfying $f_{r^*}=\log(\pi^1 / \pi^0)$ in part (ii) of
\Theorem{ChiSqMM} can be relaxed. In the case of a linear
function class we have the following extension of part (ii).

\begin{theorem} \label{t:Linearoffset}
Suppose that the observation sequence $\bfmZ$ is drawn i.i.d.
with marginal $\pi^1$ satisfying $\pi^1 \prec \pi^0$. Let
$\clF$ be the linear function class defined in
(\ref{eqn:linearfly}). Suppose the supremum in the definition
of $\DMM(\pi^1 \| \pi^0)$ is achieved at some $r^1 \in \Re^d$.
Further,  suppose that the functions $\{\psi_i\}$ satisfy the
linear independence condition of Assumption~\Ass{nondegenerate}
with $\mu^0=\pi^1$.
%
Then we have,
\begin{eqnarray*}
\lim_{n \to \infty} \Expect[n (\DMM(\Gamma^n \| \pi^0) - \DMM(\pi^1\| \pi^0))] &=& \half \trace (\Sigma_{\pi^1} \Sigma_{\cpi}^{-1}) \nonumber\\
\lim_{n \to \infty}\Var [n^\half \DMM(\Gamma^n \| \pi^0)]
&=& \sigma^2_1\\
n^\half (\DMM(\Gamma^n \| \pi^0) - \DMM(\pi^1 \| \pi^0) ) &\xrightarrow[n \to \infty]{d.}& \clN(0,\sigma^2_1)
\end{eqnarray*}
where in the first limit $\cpi = \pi^0 \exp(f_{r^1} -
\Lambda_{\pi^0}(f_{r^1}))$, and $\Sigma_{\pi^1}$ and
$\Sigma_{\cpi}$ are defined as in (\ref{e:Sigmapi}).  In the
second two limits $\sigma^2_1= \Cov_{\pi^1} (f_{r^1})$. \qed
\end{theorem}
Although we have not explicitly imposed
Assumption~\Ass{supunique} in \Theorem{Linearoffset}, the
argument we presented following \Theorem{ChiSqMM} ensures that
when $\pi^1 \prec \pi^0$, Assumption~\Ass{supunique} is
satisfied whenever Assumption~\Ass{nondegenerate} holds.
Furthermore, it can be shown that the achievement of the
supremum required in \Theorem{Linearoffset} is guaranteed if
$\pi^1$ and $\pi^0$ have equal supports. We also note that the
vector $s$ appearing in eq.~\eqref{e:psipi} of
Assumption~\Ass{nondegenerate}  is arbitrary when the
parametrization of the function class is linear.

\medskip

The weak convergence results in  \Theorem{ChiSqMM}~(i)  can be
derived from Clarke and Barron~\cite{clabar89,clabar90} (see
also \cite[Theorem 4.2]{csishi04}),   following the
maximum-likelihood estimation interpretation of the mismatched
test obtained in \Proposition{MM=ML}. In the statistics
literature, such results are called \textit{Wilks phenomena}
after the initial work by Wilks \cite{wil38}.

These results can be used to set thresholds for a target false alarm probability
in the mismatched test, just like we did for the Hoeffding test
in (\ref{e:ChiBdd}). It is shown in \cite{unnmeyvee10} that
such results can be used to set thresholds for the robust
hypothesis testing problem described in Section
\ref{s:log-linear}.

\noindent \textit{Implications for Hoeffding test} \quad The
divergence can be interpreted as a special case of mismatched
divergence defined with respect to a linear function class.
Using this interpretation, the results of \Theorem{ChiSqMM} can
also be specialized to obtain results on the Hoeffding test
statistic. To satisfy the uniqueness condition of
Assumption~\Ass{supunique}, we require that the function class
should not contain any constant functions. Now suppose that the
span of the linear function class $\clF$ together with the
constant function $f^0\equiv 1$ spans the set of all functions
on $\zstate$. This together with Assumption~\Ass{nondegenerate}
would imply that  $d = N-1$, where $N$ is the size of the
alphabet $\zstate$.  It follows from \Proposition{DivRep} that
for such a function class the mismatched divergence coincides
with the divergence. Thus, an application of
\Theorem{ChiSqMM}~(i) gives rise to the results stated in
\Theorem{biasvar1}.

\medskip

To prove \Theorem{ChiSqMM} and \Theorem{Linearoffset} we need
the following lemmas, whose proofs are given in the Appendix.

The following lemma will be used to deduce part (ii) of
\Theorem{ChiSqMM} from part (i).
\begin{lemma}
\label{t:emmdecomp} Let $\DMM_{\clF}$ denote the mismatched
divergence defined using function class $\clF$. Suppose $\pi^1
\prec \pi^0$ and the supremum in the definition of
$\DMM_{\clF}(\pi^1 \| \pi^0)$ is achieved at some $f_{r^*}\in
\clF$. Let $\cpi = \pi^0 \exp(f_{r^*} -
\Lambda_{\pi^0}(f_{r^*}))$ and $\clG=\clF-f_{r^*} \eqdef
\{f_r-f_{r^*}: r \in \Re^d \}$. Then for any $\mu$ satisfying
$\mu \prec \pi^0$, we have
\begin{equation} \DMM_{\clF}(\mu \|
\pi^0)=\DMM_{\clF}(\pi^1\|\pi^0)+\DMM_{\clG}(\mu \| \cpi)+
\langle\mu-\pi^1,\log(\frac{\cpi}{\pi^0})\rangle.\label{e:dmmdecom}\end{equation}
\qed
\end{lemma}

Suppose we apply the decomposition result from
\Lemma{emmdecomp} to the type of the observation sequence
$\bfmZ$, assumed to be drawn i.i.d.\, with marginal $\pi^1$. If
there exists $r^*$ satisfying $f_{r^*}= \log(\pi^1 / \pi^0)$,
then we have $\cpi=\pi^1$. The decomposition becomes
\begin{equation}
\DMM_{\clF}(\Gamma^n \|  \pi^0)
=
\DMM_{\clF}(\pi^1\|\pi^0)+\DMM_{\clG}(\Gamma^n \| \pi^1)
+
\langle\Gamma^n-\pi^1  ,  f_{r^*} \rangle.  
\label{e:dmmdecom2}
\end{equation}
For large $n$, the second term in the decomposition
\eqref{e:dmmdecom2} has a mean of order $n^{-1}$ and variance
of order $n^{-2}$, as shown in part (i) of  \Theorem{ChiSqMM}.
The third term has zero mean and variance of order $n^{-1}$,
since by the Central Limit Theorem,
\begin{equation}
n^\half\langle\Gamma^n-\pi^1,f_{r^*}\rangle
\xrightarrow[n \to \infty]{d.} \clN (0,\Cov_{\pi^1} (f_{r^*})
).\label{e:thirdtermdis}
\end{equation}
Thus, the asymptotic variance of $\DMM_{\clF}(\Gamma^n \|
\pi^0)$ is dominated by that of the third term and the
asymptotic bias is dominated by that of the second term. Thus
we see that part (ii) of \Theorem{ChiSqMM} can be deduced from
part (i).

\begin{lemma}
\label{t:newllnboth} Let $\bfmX = \{X^i: i = 1,2,\ldots \}$ be
an i.i.d.\  sequence with mean $\barx$ taking values in a
compact convex set $\state \subset \Re^m $, containing $\barx $
as a relative interior point. Define $S^n = \frac{1}{n}\sum_{i
= 1}^n X^i$. Suppose we are given a function $h: \Re^m \mapsto
\Re$, together with a compact set $K$ containing $\barx$ as a relative interior point such that,
\begin{enumerate}
\item The gradient $\nabla h(x)$ and the Hessian $\nabla^2
    h(x)$ are continuous over a neighborhood of $K$.
\item
$\displaystyle
\lim_{n \to \infty} -\frac{1}{n} \log \mathsf{P} \{S^n
    \notin K \} >
    0$.
\end{enumerate}
Let $M = \nabla^2 h(\barx)$ and $\Xi = \Cov(X^1)$. Then,
\begin{romannum}
\item The normalized asymptotic bias of $\{h(S^n)
    : n\ge 1\}$ is obtained via,
\[
\begin{aligned}
\lim_{n \to \infty} n\Expect[h(S^n) - h(\barx)]
&=
    \half \trace(M \Xi)
\end{aligned}
\]
\item If in addition to the above conditions, the directional
    derivative satisfies $\nabla h(\barx)^\transpose (X^1 -\barx) = 0 $
    almost surely, then the asymptotic variance decays as $n^{-2}$,
    \notes{8-16 -- have we explained these assumptions?}
    with
 \[
\lim_{n \to \infty} n^2\Var[ h(S^n)] = \half \trace(M
    \Xi M \Xi)
\]
\end{romannum}
\qed
\end{lemma}

\begin{lemma} \label{t:cltnew}
Suppose that the observation sequence $\bfmZ$ is drawn i.i.d. with
marginal $\mu \in \pz$. Let $h: \pz \mapsto \Re$ be a continuous
real-valued function whose gradient and Hessian are continuous in a
neighborhood of $\mu$. 
%
If the directional derivative
    satisfies $\nabla h(\mu)^\transpose (\nu - \mu ) \equiv 0$ for all $\nu \in
    \pz$, then
\begin{eqnarray}
n (h(\Gamma^n) - h(\mu)) \xrightarrow[n \to \infty]{d.} \half W^\transpose M W \label{eqn:clt2}
\end{eqnarray}
where $M = \nabla^2 h(\mu)$ and $W \sim \clN(0, \Sigma_W)$ with
$\Sigma_W = \mbox{diag} (\mu) - \mu \mu^\transpose$.
\qed
\end{lemma}

\begin{lemma}
\label{t:chiSq} Suppose that $V$ is an $m$-dimensional,
$\clN(0,I_m)$ random variable, and $D\colon\Re^m\to\Re^m$ is a
projection matrix.   Then $\xi \eqdef \| DV\|^2$ is a
chi-squared random variable with $K$ degrees of freedom, where
$K$ denotes the rank of $D$. \qed
\end{lemma}

\medskip

Before we proceed to the proofs of \Theorem{ChiSqMM} and
\Theorem{Linearoffset}, we recall the optimization problem
(\ref{e:MMreverse}) defining the mismatched divergence:
\begin{equation}
\DMM(\mu \| \pi^0)=  \sup_{r \in \Re^d}  \bigl(\mu(f_r) - \Lambda_{\pi^0}(f_r) \bigr).
\label{e:DMMdefn}
\end{equation}
The first order condition for optimality is given by,
\begin{equation}
g(\mu,r)=0\label{e:firstorder}
\end{equation}
where $g$ is the vector valued function that defines the gradient of the objective function in \eqref{e:DMMdefn}:
\begin{equation}
\begin{aligned}
g(\mu,r) &\eqdef \nabla_r
 \bigl(\mu(f_r) - \Lambda_{\pi^0}(f_r) \bigr)
\\
&=
\mu(\nabla_r f_{r})-\frac{\pi^0(e^{f_{r}}\nabla_r f_r)}{\pi^0(e^{f_{r}})}
\label{e:defg}
\end{aligned}
\end{equation}
The derivative of $g(\mu,r)$ with respect to $r$ is given by
\begin{eqnarray}
{\nabla_r g(\mu,r)}
&=&\mu(\nabla_r^2 f_{r})
-\left[\frac{\pi^0\bigl(e^{f_{r}}
\nabla_r f_{r}\nabla_r f_{r}^\transpose
\bigr)+\pi^0\bigl(e^{f_{r}}\nabla_r^2
f_{r}\bigr)}{\pi^0\bigl(e^{f_{r}}\bigr)}\right.\nonumber\\
&&\qquad \qquad \left.-\frac{\pi^0\bigl(e^{f_{
r}}\nabla_r f_{r}\bigr)\pi^0\bigl(e^{f_{
r}}\nabla_r
f_{r}^\transpose\bigr)}{(\pi^0\bigl(e^{f_{r}}\bigr))^2} \right]\label{e:secondfirstorder}
\end{eqnarray}
In these formulae we have extended the definition of $\mu(M)$
for  matrix-valued functions $M$ on $\zstate$ via
$[\mu(M)]_{ij}\eqdef \mu(M_{ij}) =\sum_z M_{ij}(z)\mu(z)$.
On letting $\psi^r = \nabla_r f_{r}$ we obtain,
\begin{eqnarray}
g(\mu,r) &=&
\mu(\psi^r)- \cpi^r(\psi^r)
\label{e:defgB}
\\
\nabla_r g(\mu,r)
&=&
\mu(\nabla_r^2 f_{r})
-\cpi^r(\nabla_r^2 f_{r})
-\nonumber\\
&& \qquad
\left[
\cpi^r(\psi^r{\psi^r}^\transpose) - \cpi^r(\psi^r)\cpi^r({\psi^r}^\transpose)
\right]
\label{e:secondfirstorderB}
\end{eqnarray}
where the definition of the twisted distribution is as given in  \eqref{eqn:twisteddistrn}:
\[
\cpi^r \eqdef \pi^0 \exp(f_r - \Lambda_{\pi^0}(f_r) ) .
\]

\begin{IEEEproof}[Proof of \Theorem{ChiSqMM}]
Without loss of generality, we assume that $\pi^0$ has full
support over $\zstate$. Suppose that the observation sequence
$\bfmZ$ is drawn i.i.d. with marginal distribution $\pi \in
\pz$. We have $\DMM(\Gamma^n \| \pi^0) \xrightarrow[n \to
\infty]{a.s.} \DMM(\pi \| \pi^0)$ by the law of large numbers.

\noindent 1) \textit{Proof of part (i):} \quad We first prove
the results concerning the bias and variance of the mismatched
test statistic.
We apply \Lemma{newllnboth} to the function $h(\mu)  \eqdef
\DMM(\mu\| \pi^0)$.
The other terms appearing in the lemma are taken to be $X^i =
(\mathbb{I}_{z_1}(Z_i), \mathbb{I}_{z_2}(Z_i), \ldots,
\mathbb{I}_{z_{N}}(Z_i))^\transpose $, $\state =
\clP(\mathsf{Z}) $, $\barx = \pi^0$, and $S^n = \Gamma^n$.
Let $\Xi = \Cov(X^1)$. It is easy to see that $\Xi =
\mbox{diag}(\pi^0) - \pi^0 \pi^{0 \transpose}$ and
$\Sigma_{\pi^0}=\Psi\Xi\Psi^\transpose$, where $\Sigma_{\pi^0}$
is defined in (\ref{e:Sigmapi}), and $\Psi$ is a $d \times N$
matrix defined by,
\begin{equation}
\Psi(i,j) = \psi_i(z_j). \label{eqn:Psidefn}
\end{equation}
This can be expressed as the concatenation of column vectors via
$\Psi = [\psi(z_1), \psi(z_2),\ldots, \psi(z_N) ]$.

We first demonstrate that
\begin{equation}
M=\nabla^2
h(\pi_0)=\Psi^\transpose(\Sigma_{\pi^0})^{-1}\Psi,\label{e:hM}
\end{equation}
and then check to make sure that the other requirements of
\Lemma{newllnboth} are satisfied. The first two conclusions of
\Theorem{ChiSqMM}~(i) will then follow from \Lemma{newllnboth},
since
\[
\trace(M\Xi)=\trace((\Sigma_{\pi^0})^{-1}\Psi\Xi\Psi
^\transpose)=\trace(I_d)=d,
\]
and similarly $\trace(M \Xi M \Xi )=\trace(I_d)=d$.

We first prove that under the assumptions of
\Theorem{ChiSqMM}~(i), there is a function $r: \pz \mapsto \Re$
that is $C^1$ in a neighborhood of $\pi^0$ such that $r(\mu)$
solves \eqref{e:DMMdefn} for $\mu$ in this neighborhood. Under
the uniqueness assumption \Ass{supunique}, the function
$r(\mu)$ coincides with the function given in \Ass{supunique}.

By the assumptions, we know that when $\mu=\pi^0$,
\eqref{e:firstorder} is satisfied by $r^*$ with $f_{r^*}\equiv
0$. It follows that $\pi^0 = \cpi^{r^*}$. Substituting this
into \eqref{e:secondfirstorderB}, we obtain $\nabla_r
g(\mu,r)\Big|_{\atop{\mu=\pi^0 }{ r=r^*}}= -\Sigma_{\pi^0}$,
which is negative-definite by Assumption~\Ass{nondegenerate}
and \Lemma{Siginv}. Therefore, by the Implicit Function
Theorem, there is an open neighborhood $U$ around $\mu=\pi^0$,
an open neighborhood $V$ of $r^*$, and a continuously
differentiable function $r:U\rightarrow V$ that satisfies
$g(\mu,r(\mu))=0$, for $\mu \in U$. This fact together with
Assumptions \Ass{supunique} and \Ass{nondegenerate} ensure that
when $\mu \in U \cap B$, the vector $r(\mu)$ uniquely achieves
the supremum in \eqref{e:DMMdefn}.
%
%
%

Taking the total derivative of \eqref{e:firstorder} with respect to
$\mu(z)$ we get,
\begin{eqnarray}
\frac{\partial r(\mu)}{\partial \mu(z)}&=&-\Bigl[\nabla_r
g(\mu,r(\mu))\Bigr]^{-1}\frac{\partial g(\mu,r(\mu))}{\partial
\mu(z)}. \label{e:pi0IMFDMMa}
\end{eqnarray}
Consequently, when $\mu=\pi^0$,
\begin{eqnarray}
\frac{\partial r(\mu)}{\partial \mu(z)} \Bigg|_{\mu =
\pi^0}=\Sigma_{\pi^0}^{-1}\psi(z). \label{e:pi0IMFDMM}
\end{eqnarray}
These results enable us to identify the first and second order
derivative of $h(\mu) = \DMM(\mu \| \pi^0)$. Applying
$g(\mu,r(\mu))=0$, we obtain the derivatives of $h$ as follows,
\begin{eqnarray}
\frac{\partial}{\partial \mu(z)}h(\mu) &=&f_{r(\mu)}(z).
\label{e:firstorderh}
\\
\frac{\partial^2}{\partial \mu(z)\partial \mu(\bar{z})}h(\mu)
    &=&  (\nabla_r f_{r(\mu)}(z))^\transpose \frac{\partial
r(\mu)}{\partial \mu(\bar{z})}.\label{e:hessianh}
\end{eqnarray}
When $\mu=\pi^0$, substituting \eqref{e:pi0IMFDMM} in
\eqref{e:hessianh}, we obtain \eqref{e:hM}.

We now verify the remaining conditions required for applying
\Lemma{newllnboth}:
\begin{alphanum}
 \item It is straightforward to see that $h(\pi^0)=0$.

 \item The function $h$ is uniformly bounded since $h(\mu) = \DMM(\mu\|\pi^0) \leq D(\mu\|\pi^0) \leq \max_z \log(\frac{1}{\pi^0(z)})$ and $\pi^0$ has full support.

 \item Since $f_{r(\mu)}=0$ when $\mu=\pi^0$,  it follows by \eqref{e:firstorderh} that $\frac{\partial}{\partial \mu(z)}h(\mu)\Big|_{\mu=\pi^0}=0$.
 \item Pick a compact $K \subset U\cap B$ so that $K$ contains $\pi^0$ as a relative interior point, and $K\subset \{\mu \in \pz: \max_u|\mu(u) - \pi^0(u)| <      \half\min_{u} |\pi^0(u)|\}$.
 This choice of $K$ ensures that $\lim_{n \to \infty}-\frac{1}{n} \log \mathsf{P} \{S^n \notin K \} > 0$. Note that since $r(\mu)$ is continuously differentiable on $U \cap B$, it follows by \eqref{e:firstorderh} and \eqref{e:hessianh} that $h$ is $C^2$    on      $K$.
\end{alphanum}
Thus the results on convergence of the bias and variance follow from
\Lemma{newllnboth}.

\medskip

The weak convergence result is proved using \Lemma{cltnew} and
\Lemma{chiSq}. We observe that the covariance matrix of the
Gaussian vector $W$ given in \Lemma{cltnew}  is $\Sigma_W =\Xi
= \mbox{diag}(\pi^0) -\pi^0 \pi^{0\transpose}$. This  does not
have full rank since $\Xi \mathbf{1} = 0$, where $\mathbf{1}$
is the $N\times 1$ vector of ones. Hence we can write,
\[
\Xi = G G^\transpose
\]
where $G$ is an $N \times k$ matrix for some $k < N$. In fact,
since the support of $\pi^0$ is full, we have $k=N-1$ (see
\Lemma{Siginv}). Based on this representation we can write $W =
G V$, where $V \sim \clN(0, I_{k})$.

Now, by \Lemma{cltnew}, the limiting random variable is given
by $U \eqdef \half W^\transpose M W = \half V^\transpose
G^\transpose M G V$, where $M = \nabla^2_\mu \DMM(\mu \|
\pi^0)\bigg|_{\pi^0} = \Psi^\transpose
(\Psi\Xi\Psi^\transpose)^{-1} \Psi$. We observe that the matrix
$D = G^\transpose M G$ satisfies $D^2 =D$. Moreover, since
$\Psi \Xi\Psi^\transpose$ has rank $d$ under
Assumption~\Ass{nondegenerate}, matrix $D$ also has rank $d$.
Applying \Lemma{chiSq} to matrix $D$, we conclude that $U \sim
\half \chi^2_{d}$.



\noindent 2) \textit{Proof of part (ii):}
The conclusion of part (ii) is derived using part (i) and the
decomposition in \eqref{e:dmmdecom2}. We will study the bias,
variance, and limiting distribution of each term in the
decomposition.

For the second term, note that the dimensionality of the function class $\clG$ is also $d$. Applying part (i) of this theorem to $\DMM_{\clG}(\Gamma^n \| \pi^1)$,
we conclude that its asymptotic bias and variance are given by
\begin{eqnarray}
\lim_{n \to \infty} \Expect[n \DMM_{\clG}(\Gamma^n \| \pi^1)] &=& \half d, \label{e:part2term2bias}\\
\lim_{n \to \infty} \Var [n \DMM_{\clG}(\Gamma^n \| \pi^1)]&=&\half
d.\label{e:part2term2var}
\end{eqnarray}
For the third term, since $\bfmZ$ is i.i.d. with marginal
$\pi^1$, we have
\begin{eqnarray}
\Expect[\langle\Gamma^n-\pi^1,f_{r^*}\rangle] &=& 0,\label{e:part2term3mean}\\
\Var[n^\half\langle\Gamma^n-\pi^1,f_{r^*}\rangle]&=&\Cov_{\pi^1} (f_{r^*}).\label{e:part2term3var}
\end{eqnarray}
%
The bias result \eqref{e:part2bias} follows by combining
\eqref{e:part2term2bias}, \eqref{e:part2term3mean} and using
the decomposition \eqref{e:dmmdecom2}. To prove the variance
result \eqref{e:part2var}, we again apply the decomposition
\eqref{e:dmmdecom2} to obtain,
\begin{eqnarray}
{\lim_{n \to \infty}\Var[n^\half \DMM_{\clF}(\Gamma^n \|
\pi^0)]} &=&\lim_{n \to \infty} \left\{\Var[n^\half \DMM_{\clG}(\Gamma^n \|
\pi^1)]+ \Var[n^\half
\langle\Gamma^n-\pi^1,f_{r^*}\rangle]\nonumber \right.\\&&\left.+ 2E\left[n^\half \bigl(\DMM_{\clG}(\Gamma^n \|
\pi^1)-\Expect[\DMM_{\clG}(\Gamma^n \|
\pi^1)]\bigr)\right. \right.\nonumber\\
&&\left. \left. \qquad \qquad \qquad \qquad \qquad n^\half\langle\Gamma^n-\pi^1,f_{r^*}\rangle\right] \right\}.\label{e:combvar}
\end{eqnarray}
From \eqref{e:part2term2var} it follows that the limiting value
of the first term on the right hand side of \eqref{e:combvar}
is $0$. The limiting value of the third term is also $0$ by
applying the Cauchy-–Bunyakovsky--Schwarz inequality. Thus,
\eqref{e:combvar} together with \eqref{e:part2term3var} gives
\eqref{e:part2var}.

Finally, we prove the weak convergence result
\eqref{e:part2dis} by again applying the decomposition
\eqref{e:dmmdecom2}. By \eqref{e:part2term2bias} and
\eqref{e:part2term2var}, we conclude that the second term
$n^{\frac{1}{2}} \DMM_{\clG}(\Gamma^n \| \pi^1)$ converges in
mean square to $0$ as $n \rightarrow \infty$. The weak
convergence of the third term is given in
\eqref{e:thirdtermdis}. Applying Slutsky's theorem, we obtain
\eqref{e:part2dis}.
\end{IEEEproof}

\begin{IEEEproof}[Proof of \Theorem{Linearoffset}]
The proof of this result is very similar to that of
\Theorem{ChiSqMM}~(ii) except that we use the decomposition in
\eqref{e:dmmdecom} with $\mu=\Gamma^n$. We first prove the
following generalizations of \eqref{e:part2term2bias} and
\eqref{e:part2term2var} that characterizes the asymptotic mean
and variance of the second term in \eqref{e:dmmdecom} with
$\mu=\Gamma^n$:
\begin{eqnarray}
\lim_{n \to \infty} \Expect[n \DMM_{\clG}(\Gamma^n \| \cpi)] &=& \half \trace \bigl(\Sigma_{\pi^1} (\Sigma_{\cpi})^{-1}\bigr) \label{e:proplinearterm2bias}\\
{\lim_{n \to \infty} \Var [n \DMM_{\clG}(\Gamma^n \| \cpi)] } &=&\half
\trace \left(\Sigma_{\pi^1} (\Sigma_{\cpi})^{-1} 
\Sigma_{\pi^1}
(\Sigma_{\cpi})^{-1}\right)\label{e:proplinearterm2var}
\end{eqnarray}
where $\clG=\clF-f_{r^1}$, and $\cpi$ is defined in the
statement of the proposition. The argument is similar to that
of \Theorem{ChiSqMM}~(i): We denote $\tilde{f}_{r}\eqdef
f_r-f_{r^1}$, and define
$h(\mu)\eqdef\DMM_{\clG}(\mu\|\cpi)=\sup_{r \in \Re^d}
\bigl(\mu(\tilde{f}_r) - \Lambda_{\cpi}(\tilde{f}_r) \bigr)$.
To apply \Lemma{newllnboth}, we prove the following
\begin{eqnarray}
h(\pi^1) &=&0, \label{e:hcpi0}
\\
\nabla_\mu h(\pi^1) &=& 0, \label{e:hcpifirstorder}
\\
\hbox{\it and} \qquad M = \nabla^2_\mu h(\pi^1) &=& \Psi^\transpose
(\Sigma_{\cpi})^{-1} \Psi. \label{e:hcpisecondorder}
\end{eqnarray}
The last two inequalities \eqref{e:hcpifirstorder} and
\eqref{e:hcpisecondorder} are analogous to
\eqref{e:firstorderh} and \eqref{e:hessianh}. We can also
verify that the rest of the conditions of \Lemma{newllnboth}
hold. This establishes \eqref{e:proplinearterm2bias} and
\eqref{e:proplinearterm2var}.

To prove \eqref{e:hcpi0}, first note that the supremum in the
optimization problem defining $\DMM(\pi^1\|\cpi)$ is achieved
by $\tilde{f}_{r^1}$, and we know by definition that
$\tilde{f}_{r^1}=0$. Together with the definition
$\DMM(\pi^1\|\cpi)=\pi^1(\tilde{f}_{r^1}) -
\Lambda_{\cpi}(\tilde{f}_r)$, we obtain \eqref{e:hcpi0}.

Redefine $g(\mu,r) \eqdef \nabla_r \bigl(\mu(\tilde{f}_{r}) -
\Lambda_{\cpi}(\tilde{f}_{r}) \bigr)$. The first order
optimality condition of the optimization problem defining
$\DMM(\mu\|\cpi)$ gives $g(\mu,r)=0$. The assumption that
$\clF$ is a linear function class implies that $\tilde{f}_r$ is
linear in $r$. Consequently $\nabla_r^2 \tilde{f}_{r}=0$. By
the same argument that leads to \eqref{e:secondfirstorder}, we
can show that
\begin{eqnarray}
\nabla_r g(\mu,r) &=& -\left[\frac{\cpi\bigl(e^{\tilde{f}_{r}}
\nabla_r \tilde{f}_{r}\nabla_r \tilde{f}_{r}^\transpose
\bigr)}{\cpi\bigl(e^{\tilde{f}_{r}}\bigr)}-  \frac{\cpi\bigl(e^{\tilde{f}_{
r}}\nabla_r \tilde{f}_{r}\bigr)\cpi\bigl(e^{\tilde{f}_{ r}}\nabla_r
\tilde{f}_{r}^\transpose\bigr)}{(\cpi\bigl(e^{\tilde{f}_{r}}\bigr))^2}
\right]\label{e:nableg2}
\end{eqnarray}
Together with the fact that $\tilde{f}_{r^1}=0$ and $\nabla_r
\tilde{f}_{r} = \nabla_r f_{r}$, we obtain
\begin{equation} \nabla_r g(\mu,r)\Big|_{\atop{\mu=\pi^1 }{
r=r^1}}=-\Sigma_{\cpi}.\label{e:nablag1}
\end{equation} Proceeding as in the proof of \Theorem{ChiSqMM}~(i), we
obtain \eqref{e:hcpifirstorder} and \eqref{e:hcpisecondorder}.

%


Now using similar steps as in the proof of
\Theorem{ChiSqMM}~(ii), and noticing that
$\log(\frac{\cpi}{\pi^0})=f_{r^1}$, we can establish the
following results on the third term of (\ref{e:dmmdecom}):
\begin{eqnarray*}
\Expect[\langle\Gamma^n-\pi^1,\log(\frac{\cpi}{\pi^0})\rangle]&=&0\\
\Var[n^\half\langle\Gamma^n-\pi^1,\log(\frac{\cpi}{\pi^0})\rangle]&=&\Cov_{\pi^1} (f_{r^1})\\
n^\half\langle\Gamma^n-\pi^1,\log(\frac{\cpi}{\pi^0})\rangle
&\xrightarrow[n \to \infty]{d.}& \clN (0,\Cov_{\pi^1}
(f_{r^1})).
\end{eqnarray*}
Continuing the same arguments as in \Theorem{ChiSqMM}~(i), we
obtain the result of \Theorem{Linearoffset}.
\end{IEEEproof}
\medskip
\subsection{Interpretation of the asymptotic results and performance
comparison}
The asymptotic results established above can be used to study
the finite sample performance of the mismatched test and
Hoeffding test.  Recall that in the discussion surrounding
\Figure{errLovely} we concluded that the approximation obtained
from a Central Limit Theorem gives much better estimates of
error probabilities as compared to those suggested by Sanov's
theorem.

\notes{s: minor changes throughout this section
\\
I moved the figure here as well (to generate curiosity)}

\begin{figure}[h]
\Ebox{.7}{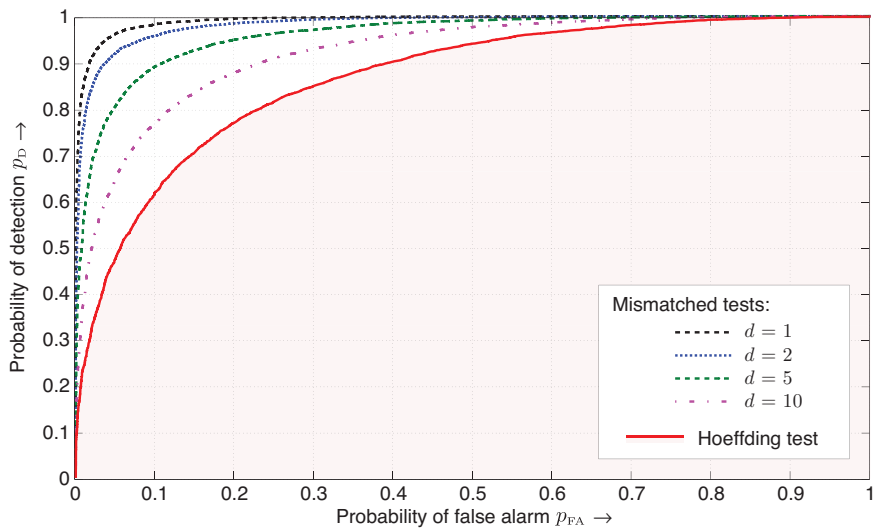}
\caption{\textit{Comparison of ROCs of Hoeffding and mismatched tests.}}
\label{f:ROCs}
\end{figure}

Suppose the log-likelihood ratio {function} $\log (\pi^1/
\pi^0)$ lies in the function class $\clF$. In this case, the
results of \Theorem{ChiSqMM} and \Lemma{emmdecomp} are
informally summarized in the following approximations: With
$\Gamma^n$ denoting the empirical distributions of the i.i.d.\
process $\bfmZ$,
 \begin{eqnarray}
{ \DMM(\Gamma^n \| \pi^0)} \approx
 \begin{cases}
D(\pi^0\| \pi^0)
 + \half \frac{1}{n}  \sum_{k=1}^d W_k^2\, , & Z_i\sim \pi^0
 \\
 D(\pi^1\| \pi^0)
 + \half \frac{1}{n}  \sum_{k=1}^d W_k^2
 +
 \frac{1}{\sqrt{n}}  \sigma_1 U\, , & Z_i\sim \pi^1
 \end{cases}
 \nonumber\\
\label{e:MMapprox}
\end{eqnarray}
where $\{ W_k\}$ is i.i.d., $N(0,1)$, and $U$ is also $N(0,1)$
but not independent of the $W_k$'s. The standard deviation
$\sigma_1$ is given in \Theorem{ChiSqMM}. These distributional
approximations are valid for large $n$, and are subject to
assumptions on the function class used in the theorem.

We observe from (\ref{e:MMapprox}) that, for large enough $n$,
when the observations are drawn under $\pi^0$, the mismatched
divergence is well approximated by $\frac{1}{2n}$ times a
chi-squared random variable with $d$ degrees of freedom. We
also observe that when the observations are drawn under
$\pi^1$, the mismatched divergence is well approximated by a
Gaussian random variable with mean $D(\pi^1 \| \pi^0)$ and with
a variance proportional to $\frac{1}{n}$ and independent of
$d$.
Since the mismatched test can be interpreted as a GLRT, these
results capture the rate of degradation of the finite sample
performance of a GLRT as the dimensionality of the
parameterized family of alternate hypotheses increases. We
corroborate this intuitive reasoning through Monte Carlo
simulation experiments.

We estimated via simulation the performance of the Hoeffding
test and mismatched tests designed using a linear function
class. We compared the error probabilities of these tests for
an alphabet size of $N = 19$ and sequence length of $n = 40$.
We chose $\pi^0$ to be the uniform distribution, and $\pi^1$ to
be the distribution obtained by convolving two uniform
distributions on sets of size $(N+1)/2$. We chose the basis
function $\psi_1$ appearing in (\ref{eqn:linearfly}) to be the
log-likelihood ratio between $\pi^1$ and $\pi^0$, viz.,
\[
\psi_1(z_i) = \log \frac{\pi^1(z_i)}{\pi^0(z_i)}, \quad 1 \leq i \leq N
\]
and the other basis functions $\psi_2, \psi_3, \ldots,
\psi_{d}$ were chosen uniformly at random. \Figure{ROCs} shows
a comparison of the ROCs of the Hoeffding test and mismatched
tests for different values of dimension $d$.
Plotted on the $x$-axis is the probability of false alarm, i.e., the probability of
misclassification under $\pi^0$;  shown on the $y$-axis is the
probability of detection, i.e., the probability of correct
classification under $\pi^1$. The various points on each ROC
curve are obtained by varying the threshold $\eta$ used in the
Hoeffding test of (\ref{eqn:Hoefftestnew2}) and mismatched test
of (\ref{eqn:MMtest}).

From \Figure{ROCs} we see that as $d$ increases the performance
of the mismatched tests degrades. This is consistent with the
approximation \eqref{e:MMapprox}
which suggests that the variance of the mismatched divergence
increases with $d$. Furthermore, as we saw earlier, the
Hoeffding test can be interpreted as a special case of the
mismatched test for a specific choice of the function class
with $d = N-1$ and hence the performance of the mismatched test
matches the performance of the Hoeffding test when $d = N-1$.


To summarize, the above results suggest that although the
Hoeffding test is optimal in an error-exponent sense, it is
disadvantageous in terms of finite sample error probabilities
to blindly use the Hoeffding test if it is known a priori that
the alternate distribution belongs to some parameterized family
of distributions.

\section{Conclusions}
\label{s:conc}

The mismatched test provides a solution to the universal
hypothesis testing problem that can incorporate prior knowledge
in order to reduce variance.   The main results of
\Section{s:var} show that the variance reduction over
Hoeffding's optimal test is substantial when the state space is
large.

The dimensionality of the function class can be chosen by the
designer to ensure that the the bias and variance are within
tolerable limits.  It is in this phase of design that prior
knowledge is required to ensure that the  error-exponent
remains sufficiently large under the alternate hypothesis (see
e.g.\ \Corollary{MMoptwithLLR}).  In this way the designer can
make effective tradeoffs between the power of the test and the
variance of the test statistic.


The mismatched divergence provides a unification of several
approaches to robust and universal hypothesis testing. Although
constructed in an i.i.d.\ setting, the mismatched tests are
applicable in very general settings,  and the performance
analysis presented here is easily generalized to any stationary
process satisfying the Central Limit Theorem.

There are many directions for future research.  Topics of current research include,
\begin{romannum}

\item
Algorithms for basis synthesis and basis adaptation.

\item
Extensions to Markovian models.

\item
Extensions to change detection.
\end{romannum}
Initial progress in basis synthesis is reported in
\cite{huamey10}. Recent results addressing the computational
complexity of the mismatched test are reported in \cite{unn10}.
Although the exact computation of the mismatched divergence
requires the solution of an optimization problem, we describe a
computationally tractable approximation in \cite{unn10}. We are
also actively pursuing applications to problems surrounding
building energy and surveillance.  Some initial progress is
reported in \cite{meysurlinnar09a}.

\section*{Acknowledgements}
We thank Prof. Imre Csisz\'{a}r for his insightful comments.

\appendix
\section{Appendix}

\subsection{Excess codelength for source coding with training}
The results in Theorem~\ref{t:ChiSqMM}  give us the asymptotic
behavior of $ D(\Gamma^n \| \pi)$ but what we need here is the
behavior of $D(\pi\| \Gamma^n )$. Define
\[
h(\mu) =\left\{ \begin{tabular}{cc}
$D(\pi \| \mu)$& if $\mu \in \bP_{\epsilon/2}$\\
$D(\pi \| \pi^\uni)$ &else
\end{tabular} \right. .
\]
It is clear that $h$ is uniformly bounded from above by $\log
\frac{2}{\epsilon}$. Although $h$ is not continuous at the
boundary of $\bP_{\epsilon/2}$, a modified version of Lemmas
\ref{t:newllnboth} and \ref{t:cltnew} can be applied to the
function $h$ to establish the results of (\ref{eqn:srccoding})
following the same steps used in proving Theorem
\ref{t:ChiSqMM}. The Hessian matrix $M$ appearing in the
statement of the lemmas is given by,
\[
M = \nabla^2 h(\pi) = \mbox{ diag} (\pi)^{-1}.
\]
Hence, $\mbox{trace}(M\Omega) = \mbox{trace}(M\Omega M\Omega) =
N-1$.

\subsection{Proof of \Lemma{emmdecomp}}
\begin{IEEEproof}
In the following chain of identities, the first, third and
fifth equalities follow from relation (\ref{e:MMreverse}) and
\Proposition{Pythag}.
\begin{eqnarray}
{\DMM_{\clF}(\mu \| \pi^0)} &=&D(\mu\|\pi^0) -\inf\{D(\mu
\|\nu):\nu=\pi^0 \exp(f-\Lambda_{\pi^0}(f)),f\in \clF\}
\nonumber\\
&=&D(\mu\|\cpi)+\langle\mu,\log(\frac{\cpi}{\pi^0})\rangle -\inf\{D(\mu
\|\nu):\nu=\cpi \exp(f-\Lambda_{\cpi}(f)),f\in \clG\}
\nonumber\\
&=&\DMM_{\clG}(\mu \| \cpi)+
\langle\mu,\log(\frac{\cpi}{\pi^0})\rangle
\nonumber\\
&=&\DMM_{\clG}(\mu \| \cpi)+
\langle\mu-\pi^1,\log(\frac{\cpi}{\pi^0})\rangle+D(\pi^1\|\pi^0)-D(\pi^1\|\cpi)
\nonumber\\
&=&\DMM_{\clG}(\mu \| \cpi)+
\langle\mu-\pi^1,\log(\frac{\cpi}{\pi^0})\rangle+\DMM_{\clF}(\pi^1\|\pi^0)\nonumber
\end{eqnarray}
\end{IEEEproof}

\subsection{Proof of \Lemma{newllnboth}}

The following simple lemma will be used in multiple places in
the proof that follows.
\begin{lemma}
\label{t:expctnprodconv}
If a sequence of random variables $\{A^n\}$ satisfies
$\Expect[A^n] \xrightarrow[n \to \infty]{} a$ and
$\{\Expect[(A^n)^2]\}$ is a bounded sequence, and another
sequence of random variables $\{B^n\}$ satisfies $B^n
\xrightarrow[n \to \infty]{m.s.} b$, then $\Expect[A^nB^n]
\xrightarrow[n \to \infty]{} ab$.
\qed
\end{lemma}

\begin{IEEEproof}[Proof of \Lemma{newllnboth}]
Without loss of generality, we can assume that the mean $\barx$
is the origin in $\Re^m$ and that $h(\barx) = 0$.

Since the Hessian is continuous over the set $K$, we have by
Taylor's theorem:
\begin{eqnarray}
{n (h(S^n) - \nabla h(\barx)^\transpose S^n )\mathbb{I}_{\{S^n \in K\}}}
&=& n[h(\barx) +  \half {S^n}^\transpose \nabla^2 h(\tilS^n) S^n]\mathbb{I}_{\{S^n \in K\}} \\
&=& \frac{n}{2} {S^n}^\transpose \nabla^2 h(\tilS^n) S^n \mathbb{I}_{\{S^n \in K\}} \label{taylor}
\end{eqnarray}
where $\tilS^n = \gamma S^n$ for some $\gamma=\gamma(n) \in [0,1]$.
By the strong law of large numbers we have $S^n \xrightarrow[n
\to \infty]{a.s.} \barx$. Hence $\tilS^n \xrightarrow[n \to
\infty]{a.s.} \barx$ and $\nabla^2 h(\tilS^n) \xrightarrow[n \to
\infty]{a.s.} \nabla^2 h(\barx) = M$ since $\nabla^2 h$ is
continuous at $\barx$. Now by the boundedness of the second
derivative over $K$ and the fact that
\[
\mathbb{I}_{\{S^n \in K \}} \xrightarrow[n \to \infty]{a.s.}1
\]
we have $(\nabla^2 h(\tilS^n))_{i,j} \mathbb{I}_{\{S^n \in K
\}} \xrightarrow[n \to \infty]{m.s.} M_{i,j}$.

Under the assumption that $\bfmX$ is i.i.d. on the compact set
$\state$, we have
\[
\Expect[n S^n_i S^n_j] = \Sigma_{i,j} \mbox{ for all }n,
\]
and $\Expect[(n S^n_i S^n_j)^2]$ converges to a finite quantity
as $n \to \infty$. Hence the results of Lemma
\ref{t:expctnprodconv} are applicable with $A^n = n S^n_i
S^n_j$  and $B^n = \nabla^2 h(\tilS^n)_{i,j}\mathbb{I}_{\{S^n
\in K \}}$, which gives:
\begin{equation}
\Expect[n S^n_i S^n_j \nabla^2 h(\tilS^n)_{i,j}\mathbb{I}_{\{S^n \in K \}}] \xrightarrow[n \to \infty]{}  \Sigma_{i,j} M_{i,j}.
\end{equation}
Thus we have
\begin{eqnarray}
{\Expect[n (h(S^n) - \nabla h(\barx)^\transpose S^n)\mathbb{I}_{\{S^n \in K \}}]}
 &=&\Expect[\frac{n}{2} {S^n}^\transpose \nabla^2 h(\tilS^n) S^n  \mathbb{I}_{\{S^n \in K \}}]\nonumber\\ &\xrightarrow[n \to \infty]{}& \half \text{trace}(M \Xi). \label{cgneceK}
\end{eqnarray}
Since $\state$ is compact, $h$ is continuous, and $h$ is
differentiable at $\barx$, it follows that there are scalars
$\overline{h}$ and $\overline{x}$ such that $\sup_{x \in
\state} |h(x)| \leq \overline{h}$ and $|\nabla
h(\barx)^\transpose S^n| < \overline{x}$. Hence,
\begin{eqnarray}
{|\Expect[n (h(S^n)-\nabla h(\barx)^\transpose S^n)\mathbb{I}_{\{S^n \notin K \}}]|} \leq n (\overline{h}+\overline{x}) \mathsf{P}{\{S^n \notin K \}} \xrightarrow[n \to \infty]{} 0  \label{cgneceKc}
\end{eqnarray}
where we use the assumption that the $\mathsf{P}{\{S^n \notin K
\}}$ decays exponentially in $n$. Combining (\ref{cgneceK}) and
(\ref{cgneceKc}) and using the fact that $S^n$ has zero mean,
we have
\[
\Expect[n h(S^n)]  = \Expect[n (h(S^n) - \nabla h(\barx)^\transpose S^n)] \xrightarrow[n \to \infty]{} \half \text{trace}(M \Xi).
\]
This establishes the result of (i).

Under the condition that the directional derivative is zero,
(\ref{taylor}) can be written as
\begin{eqnarray}
nh(S^n)\mathbb{I}_{\{S^n \in K\}} = \frac{n}{2} {S^n}^\transpose \nabla^2 h(\tilS^n) S^n \mathbb{I}_{\{S^n \in K\}}. \label{taylor2}
\end{eqnarray}

Now by squaring (\ref{taylor2}), we have
\begin{eqnarray*}
{(n h(S^n)\mathbb{I}_{\{S^n \in K\}})^2}
= \frac{n^2}{4} \sum_{i,j,k,\ell} \left[S^n_i (\nabla^2 h(\tilS^n))_{i,j} S^n_j S^n_k (\nabla^2 h(\tilS^n))_{k,\ell} S^n_\ell\mathbb{I}_{\{S^n \in K\}}\right].
\end{eqnarray*}
As before, by the boundedness of the Hessian we have:
\[
(\nabla^2 h(\tilS^n))_{i,j} (\nabla^2
h(\tilS^n))_{k,\ell} \mathbb{I}_{\{S^n \in K \}}
\xrightarrow[n \to \infty]{m.s.} M_{i,j}M_{k,\ell}
\]
It can also be shown that
\begin{eqnarray*}
{\Expect[n^2 S^n_i S^n_j S^n_k S^n_\ell]}=
\frac{F_{i,j,k,l}}{n}+ \Sigma_{i,j} \Sigma_{k,\ell}+ \Sigma_{j,k}
\Sigma_{i,\ell} +\Sigma_{i,k}\Sigma_{j,\ell} \mbox{ for all }n
\end{eqnarray*}
where $F_{i,j,k,l} = \Expect [X^1_i X^1_j X^1_k X^1_\ell]$.
Moreover, $\Expect[(n^2 S^n_i S^n_j S^n_k S^n_\ell)^2]$ is
finite for each $n$ and converges to a finite quantity as $n \to
\infty$ since the moments of $X^i$ are finite. Thus we can
again apply Lemma \ref{t:expctnprodconv} to see that
\begin{eqnarray*}
\lefteqn{\Expect[n^2 S^n_i \nabla^2 h(\tilS^n)_{i,j}  S^n_j S^n_k \nabla^2 h(\tilS^n)_{k,\ell}  S^n_\ell \mathbb{I}_{\{S^n \in K \}}]}\\
&\xrightarrow[n \to \infty]{}&  (\Sigma_{i,j} \Sigma_{k,\ell}+ \Sigma_{j,k} \Sigma_{i,\ell} +\Sigma_{i,k}\Sigma_{j,\ell})M_{i,j}M_{k,\ell}.
\end{eqnarray*}
Putting together terms and using (\ref{taylor2}) we obtain:
\begin{eqnarray*}
{\Expect [(n h(S^n))^2 \mathbb{I}_{\{S^n \in K\}}]}
\xrightarrow[n \to \infty]{} \half \text{trace}(M \Xi M \Xi) +
\fourth(\text{trace}(M \Xi))^2.
\end{eqnarray*}
Now similar to
(\ref{cgneceKc}) we have:
\begin{equation}
|\Expect[(n h(S^n))^2\mathbb{I}_{\{S^n
\notin K \}}]| \leq n^2 \overline{h}^2 \mathsf{P}{\{S^n \notin K \}}
\xrightarrow[n \to \infty]{} 0. \label{ldpbound}
\end{equation}
Consequently
\[
\Expect [(n h(S^n))^2 ]  \xrightarrow[n \to \infty]{} \half \text{trace}(M \Xi M \Xi) + \fourth(\text{trace}(M \Xi))^2
\]
which gives (ii).

\end{IEEEproof}

\subsection{Proof of \Lemma{cltnew}}
We know from (\ref{eqn:typeiid}) that $\Gamma^n$
can be written as an empirical average of i.i.d. vectors.
Hence, it satisfies the central limit theorem which says that,
\begin{equation}
n^\half(\Gamma^n - \mu) \xrightarrow[n \to \infty]{d.} W \label{eqn:cltGamman}
\end{equation}
where the distribution of $W$ is defined below (\ref{eqn:clt2}).

Considering a second-order Taylor's expansion and using the
condition on the directional derivative, we have,
\begin{eqnarray*}
n (h(\Gamma^n) - h(\mu)) = \half n((\Gamma^n - \mu)^\transpose  \nabla^2 h(\tilde \Gamma^n) (\Gamma^n - \mu))
\end{eqnarray*}
where $\tilde \Gamma^n  = \gamma \Gamma^n + (1 - \gamma) \mu $ for
some $\gamma= \gamma(n) \in [0,1]$. We also know by the strong law
of large numbers that $\Gamma^n$ and hence $\tilde \Gamma^n$
converge to $\mu$ almost surely. By the continuity of the Hessian,
we have
\begin{equation}
\nabla^2 h(\tilde \Gamma^n) \xrightarrow[n \to
\infty]{a.s.} \nabla^2 h(\mu).\label{eqn:conHessian}
\end{equation}
By applying the vector-version of Slutsky's theorem
\cite{bil68}, together with \eqref{eqn:cltGamman} and
\eqref{eqn:conHessian}, we conclude
\[
n((\Gamma^n - \mu)^\transpose  \nabla^2 h(\tilde \Gamma^n) (\Gamma^n - \mu)) \xrightarrow[n \to \infty]{d.} \half W^\transpose \nabla^2 h(\mu) W,
\]
thus establishing the lemma.

\subsection{Proof of \Lemma{chiSq}}
\begin{IEEEproof}
The assumption that $D$ is a projection matrix implies that
$D^2 = D$. Let $\{ u^1,\dots,u^m\}$ denote an orthonormal
basis, chosen so that the first $K$ vectors span the range
space of $D$. Hence $Du^i =u^i$ for $1\le i\le K$, and $Du^i
=0$ for all other $i$.

Let $U$ denote the unitary matrix whose $m$ columns are $\{
u^1,\dots,u^m\}$.   Then $\tilV = U V$ is also an $\clN(0,I_m)$
random variable,  and hence $DV$ and $D\tilV$ have the same
Gaussian distribution.

To complete the proof we demonstrate that  $ \| D\tilV\|^2  $
has a chi-squared distribution: By construction the vector
$\tilY = D\tilV$ has components given by
\[
\tilY_i =\begin{cases}  \tilV_i & 1\le i\le K\\  0& K<i\le m \end{cases}
\]
It follows that $\|\tilY\|^2 = \| D\tilV\|^2 = \tilV_1^2
+\cdots +  \tilV_K^2$ has a chi-squared distribution with $K$
degrees of freedom.
\end{IEEEproof}


\bibliographystyle{IEEEtran}

\begin{thebibliography}{10}
\providecommand{\url}[1]{#1}
\csname url@samestyle\endcsname
\providecommand{\newblock}{\relax}
\providecommand{\bibinfo}[2]{#2}
\providecommand{\BIBentrySTDinterwordspacing}{\spaceskip=0pt\relax}
\providecommand{\BIBentryALTinterwordstretchfactor}{4}
\providecommand{\BIBentryALTinterwordspacing}{\spaceskip=\fontdimen2\font plus
\BIBentryALTinterwordstretchfactor\fontdimen3\font minus
  \fontdimen4\font\relax}
\providecommand{\BIBforeignlanguage}[2]{{%
\expandafter\ifx\csname l@#1\endcsname\relax
\typeout{** WARNING: IEEEtran.bst: No hyphenation pattern has been}%
\typeout{** loaded for the language `#1'. Using the pattern for}%
\typeout{** the default language instead.}%
\else
\language=\csname l@#1\endcsname
\fi
#2}}
\providecommand{\BIBdecl}{\relax}
\BIBdecl

\bibitem{huaunnmeyveesur09}
D.~Huang, J.~Unnikrishnan, S.~Meyn, V.~Veeravalli, and A.~Surana, ``Statistical
  {SVMs} for robust detection, supervised learning, and universal
  classification,'' in \emph{Proc. of {IEEE Information Theory Workshop on
  Networking and Information Theory}}, Volos, Greece, 2009, pp. 62--66.

\bibitem{hoe65a}
W.~Hoeffding, ``Asymptotically optimal tests for multinomial distributions,''
  \emph{Ann. Math. Statist.}, vol.~36, pp. 369--408, 1965.

\bibitem{zeizivmer92}
O.~Zeitouni, J.~Ziv, and N.~Merhav, ``When is the generalized likelihood ratio
  test optimal?'' \emph{IEEE Trans. Inform. Theory}, vol.~38, no.~5, pp.
  1597--1602, 1992.

\bibitem{levmer02}
E.~Levitan and N.~Merhav, ``A competitive {Neyman-Pearson} approach to
  universal hypothesis testing with applications,'' \emph{IEEE Trans. Inform.
  Theory}, vol.~48, no.~8, pp. 2215--2229, 2002.

\bibitem{lap96}
A.~Lapidoth, ``Mismatched decoding and the multiple-access channel,''
  \emph{IEEE Trans. Inform. Theory}, vol.~42, no.~5, pp. 1439--1452, Sep 1996.

\bibitem{abbmedmeyzhe07a}
E.~Abbe, M.~Medard, S.~Meyn, and L.~Zheng, ``Finding the best mismatched
  detector for channel coding and hypothesis testing,'' \emph{Information
  Theory and Applications Workshop, 2007}, pp. 284--288, 29 2007-Feb. 2 2007.

\bibitem{csishi04}
I.~Csisz{\'a}r and P.~C. Shields, ``Information theory and statistics: A
  tutorial,'' \emph{Foundations and Trends in Communications and Information
  Theory}, vol.~1, no.~4, 2004.

\bibitem{panmeyvee04a}
C.~Pandit, S.~Meyn, and V.~Veeravalli, ``Asymptotic robust {Neyman-Pearson}
  hypothesis testing based on moment classes,'' in \emph{Proc. of IEEE
  International Symposium on Information Theory}, Chicago, 2004, p. 220.

\bibitem{panmey06a}
C.~Pandit and S.~P. Meyn, ``Worst-case large-deviations with application to
  queueing and information theory,'' \emph{Stoch. Proc. Applns.}, vol. 116,
  no.~5, pp. 724--756, May 2006.

\bibitem{donvarI-II}
M.~Donsker and S.~Varadhan, ``Asymptotic evaluation of certain {M}arkov process
  expectations for large time. {I}. {I}{I},'' \emph{Comm. Pure Appl. Math.},
  vol.~28, pp. 1--47; ibid. {\bf 28} (1975), 279--301, 1975.

\bibitem{nguwaijor08}
X.~Nguyen, M.~J. Wainwright, and M.~I. Jordan, ``Estimating divergence
  functionals and the likelihood ratio by convex risk minimization,''
  \emph{IEEE Trans. Inf. Theory}, 2010, to appear.

\bibitem{clabar89}
B.~Clarke and A.~R. Barron, ``Information theoretic asymptotics of bayes'
  methods,'' Univ. of Illinois, Department of Statistics, Tech. Rep.~26, July
  1989.

\bibitem{clabar90}
B.~S. Clarke and A.~R. Barron, ``Information-theoretic asymptotics of {B}ayes
  methods,'' \emph{IEEE Trans. Inform. Theory}, vol.~36, no.~3, pp. 453--471,
  1990.

\bibitem{wil38}
S.~S. Wilks, ``The large-sample distribution of the likelihood ratio for
  testing composite hypotheses,'' \emph{Ann. Math. Statistics}, vol.~9, pp.
  60--62, 1938.

\bibitem{ilt95}
\BIBentryALTinterwordspacing
M.~Iltis, ``Sharp asymptotics of large deviations in {$\Re^d$},'' \emph{Journal
  of Theoretical Probability}, vol.~8, no.~3, pp. 501--522, 1995. [Online].
  Available: \url{http://www.springerlink.com/content/3768273345028TR4}
\BIBentrySTDinterwordspacing

\bibitem{harremoes09}
P.~Harremo\"{e}s, ``Testing goodness-of-fit via rate distortion,'' in
  \emph{Proc. of IEEE Information Theory Workshop on Networking and Information
  Theory}, Volos, Greece, June 2009, pp. 17--21.

\bibitem{zeigut91a}
O.~Zeitouni and M.~Gutman, ``On universal hypotheses testing via large
  deviations,'' \emph{IEEE Trans. Inform. Theory}, vol.~37, no.~2, pp.
  285--290, 1991.

\bibitem{zeigut91b}
------, ``Correction to: ``{O}n universal hypotheses testing via large
  deviations'','' \emph{IEEE Trans. Inform. Theory}, vol.~37, no. 3, part 1, p.
  698, 1991.

\bibitem{zhofumar10}
Z.~Enlu, M.~C. Fu, and S.~I. Marcus, ``Solving continuous-state pomdps via
  density projection,'' \emph{IEEE Trans. Autom. Control}, vol.~55, no.~5, pp.
  1101 -- 1116, May 2010.

\bibitem{unn10}
J.~Unnikrishnan, ``Decision-making under statistical uncertainty,'' Ph.D.
  dissertation, University of Illinois at Urbana-Champaign, Urbana, IL, August
  2010.

\bibitem{unnmeyvee10}
\BIBentryALTinterwordspacing
J.~Unnikrishnan, S.~Meyn, and V.~V. Veeravalli, ``On thresholds for goodness of
  fit tests,'' presented at IEEE Information Theory Workshop, Dublin, 2010.
  [Online]. Available:
  \url{http://www.ifp.illinois.edu/$\sim$junnikr2/pdfs/ITW10.pdf}
\BIBentrySTDinterwordspacing

\bibitem{huamey10}
D.~Huang and S.~Meyn, ``Feature extraction for universal hypothesis testing via
  rank-constrained optimization,'' in \emph{Information Theory Proceedings
  (ISIT), 2010 IEEE International Symposium on}, Jun. 2010, pp. 1618 -- 1622.

\bibitem{meysurlinnar09a}
S.~Meyn, A.~Surana, Y.~Lin, and S.~Narayanan, ``Anomaly detection using
  projective {Markov} models in a distributed sensor network,'' in \emph{Proc.
  of 48th IEEE Conference on Decision and Control}, Shanghai, December 2009,
  pp. 4662--4669.

\bibitem{bil68}
P.~Billingsley, \emph{{Convergence of Probability Measures}}.\hskip 1em plus
  0.5em minus 0.4em\relax New York: John Wiley \& Sons, 1968.

\end{thebibliography}
\def\cprime{$'$}\def\cprime{$'$}

\end{document}